\documentclass[twocolumn]{aastex62}
\usepackage{url}
\usepackage{bigstrut,bigdelim,multirow}

\shorttitle{Precursors in SGRBs}
\shortauthors{Zhong et al.}

\begin{document}
\title{Precursors in Short Gamma-ray Bursts as a Possible Probe of Progenitors}

\author{Shu-Qing Zhong}
\affil{School of Astronomy and Space Science, Nanjing University, Nanjing 210093, China; sqzhong@hotmail.com, dzg@nju.edu.cn}
\affil{Key laboratory of Modern Astronomy and Astrophysics (Nanjing University), Ministry of Education, Nanjing 210093, China}
\author{Zi-Gao Dai}
\affil{School of Astronomy and Space Science, Nanjing University, Nanjing 210093, China; sqzhong@hotmail.com, dzg@nju.edu.cn}
\affil{Key laboratory of Modern Astronomy and Astrophysics (Nanjing University), Ministry of Education, Nanjing 210093, China}
\author{Ji-Gui Cheng}
\affil{Guangxi Key Laboratory for Relativistic Astrophysics, Department of Physics, Guangxi University, Nanning 530004, China}
\author{Lin Lan}
\affil{Guangxi Key Laboratory for Relativistic Astrophysics, Department of Physics, Guangxi University, Nanning 530004, China}
\author{Hai-Ming Zhang}
\affil{School of Astronomy and Space Science, Nanjing University, Nanjing 210093, China; sqzhong@hotmail.com, dzg@nju.edu.cn}
\affil{Key laboratory of Modern Astronomy and Astrophysics (Nanjing University), Ministry of Education, Nanjing 210093, China}

\begin{abstract}
We extract 18 candidate short gamma-ray bursts (SGRBs) with precursors from 660 SGRBs
observed by {\em Fermi} and {\em Swift} satellites, and carry out a comprehensive analysis on their
temporal and spectral features. We obtain the following results: (1) For a large fraction of candidates, the main
burst durations are longer than their precursor durations, comparable to their quiescent times
from the end of precursors to the beginning of their main bursts. (2) The average flux of precursors tends to increase as their main bursts brighten. (3) As seen from the distributions of hardness ratio and spectral fitting, the precursors are slightly spectrally softer with respect to the main bursts. Moreover, a significant portion of precursors and all main bursts favor a
non-thermal spectrum. (4) The precursors might be a probe of the progenitor properties
of SGRBs such as the magnetic field strength and the crustal equation of state if they arise
from some processes before mergers of binary compact objects rather than post-merger processes.
\end{abstract}


\section{Introduction}
\label{sec:introduction}
Based on the bimodal burst duration distribution ($T_{90}$), gamma-ray bursts (GRBs) can be generally
classified into long gamma-ray bursts (LGRBs) and short gamma-ray bursts (SGRBs) \citep{kou93},
which are widely thought to be associated with two types of distinguished origins:
the core collapse of massive stars \citep{woo93,pac98} and the merger of neutron star-neutron star (NS-NS) \citep{pac86,eich89}
or neutron star-black hole (NS-BH) \citep{pac91}, respectively.
Thus the prompt GRB emissions can statistically probe for the type of GRBs' progenitors.
Besides, many authors
investigated qualitatively and quantitatively the GRBs' progenitors from other observations such as their associated supernovae \citep[e.g.,][]{gal98,hjo03},
host galaxies \citep{bloom02,berger11}, afterglows \citep{klo04}, and gravitational waves \citep{abb17}.
In addition to these methods of probing the progenitor properties,
a different method that was suggested is to utilize the earlier less-intense episodes (the so-called ``precursors'') \citep{mur91,kos95,laz05,hu14,lan18} preceding to the prompt emission episodes (``main bursts'') in GRBs,
especially in SGRBs \citep{tro10,mp17,wang18}.

Observationally, in LGRBs detected by the {\em BATSE} telescope, the precursors earlier than the main
bursts with typical several tens of seconds extending up to 200\,s have typically a non-thermal power-law
spectrum and a feature spectrally softer than the main bursts \citep{laz05}. Moreover, a very large fraction of
precursors of LGRBs observed by {\em Swift} have a shorter duration than their main bursts \citep{hu14}.
Theoretically, precursors in LGRBs could be associated with relativistic fireballs that were proposed
to explain their main bursts \citep{pac86,mes01}, or with progenitor-linking
jet breakout from massive stars \citep{ram02,wm03,zhang03,lb05}. But these interpretations cannot completely meet the precursor
observations in LGRBs \citep{laz05}.

For precursors in SGRBs, \cite{tro10} carried out a systematic search in {\em Swift}
catalog and did not find substantial differences between the precursor and the main burst from the
hardness ratio comparison for five SGRBs. \cite{mp17} also made a comprehensive
search for the precursors of SGRBs detected in the SPI-ACS/INTEGRAL experiment and {\em Fermi} satellite
and found only four candidates. In any case, only a few SGRBs with precursors make it difficult to systematically study
the temporal and spectral features of precursors.
Therefore, more SGRBs with precursors are still expected to collect.

In this paper, we first carry out a complete search for SGRBs with precursors observed by {\em Swift} from
2004 November to 2019 May and {\em Fermi} from 2008 June to 2019 May, and extract available candidates by a Bayesian Blocks (BBlocks) technique in \S2.
We further statistically analyze their temporal and spectral properties, and the possible relations
among the quantities that represent the features in \S3. Finally we discuss the origin of the precursors in SGRBs and the possibility
of the precursors as a probe of the progenitors of SGRBs in \S4, and present conclusions in \S5.
Throughout, the notations $Q_n=Q/10^{n}$ in cgs units and a concordance cosmology
with parameters $H_0 = 70~{\rm km~s^{-1}~ Mpc^{-1}}$, $\Omega_{\rm M} = 0.30$, and $\Omega_{\Lambda} = 0.70$ are adopted.

\section{Data and Analysis}
\label{sec:data}
\subsection{Sample Selection with Bayesian Blocks Analysis}
\label{subsec:sample selection}

We have searched for SGRBs from {\em Swift} Burst Alert Telescope (BAT)\footnote{http://www.swift.ac.uk/archive/obs.php?burst=1}
and {\em Fermi} Gamma-Ray Burst Monitor (GBM) data\footnote{
https://heasarc.gsfc.nasa.gov/W3Browse/fermi/fermigbrst.html}. Adopting the burst duration requirement $T_{90}<2.0$ s,
we made a gallery of nearly 120 BAT SGRBs and 430 GBM SGRBs. To extend the SGRB sample, we have also collected more than
110 other possible SGRBs from the literature \citep[e.g.,][]{lu17}. The total number of SGRBs is up to 660.
Interestingly, some SGRBs either observed by both {\em Swift} BAT and {\em Fermi} GBM or observed by both {\em Fermi}
GBM and {\em Fermi} Large Area Telescope (LAT) can provide more information about themselves.

To extract those SGRBs with precursors, we have performed the BBlocks Representations \citep{sca13}
to find the optimal segmentation for the {\em Swift} and {\em Fermi} time-tagged event (TTE) data. This analysis technique
divides a series of events characterized by the photon arrival times into subintervals (blocks) of the perceptibly constant count rate,
with change points to define the edges of blocks and determine the signal amplitudes of blocks. Moreover, the false positive rate is given to $p_0=0.01$,
which decides the penalization on the likelihood and affects the number of blocks \citep[see Eq. (21) in][]{sca13}.
In addition, softwares {\em heasoft (ver. 6.24)}, {\em calibration database (CALDB)}, and {\em ScienceTools (v10r0p5)}
are needed to use for data
extraction and handling. As the {\em Swift} BAT TTE data are slightly different from those of {\em Fermi} in background substraction, the handling processes are somewhat different as follows:

(1) {\em Swift} SGRBs: their TTE data need to give the mask weighting
for masking out the noisy or defective detectors from the quality map \footnote{https://swift.gsfc.nasa.gov/analysis/threads/batmaskwtthread.html}.
We extracted the event data with time intervals [-200 s, 50 s] relative to their trigger times, and searched
the presence of weak emission signal by the BBlocks technique.
In order to inspect the signals within different energy bands, we used the BBlocks for events
in 15-50 keV, 50-150 keV, and the whole energy range 15-150 keV.

(2) {\em Fermi} SGRBs: these data are mainly obtained by the GBM instrument containing NaI sodium iodide (NaI)
detectors with energy band 8 keV - 1 MeV and bismuth germanate (BGO) scintillation detectors covering energy
range 200 keV - 40 MeV \citep{mee09}. A few events also are observed by LAT instrument with
energy range 20 MeV - 300 GeV \citep{atw09}. Similarly,
we performed BBlocks for their TTE data in the intervals [-200 s, 50 s] since their GBM triggers.
The energy bands are split into 8-50 keV, 50-1000 keV, and 8-1000 keV for the NaI data,
while 250-1000 keV and $>$1000 keV for those of BGO are adopted.

\begin{figure}
\includegraphics[angle=0,width=0.5\textwidth]{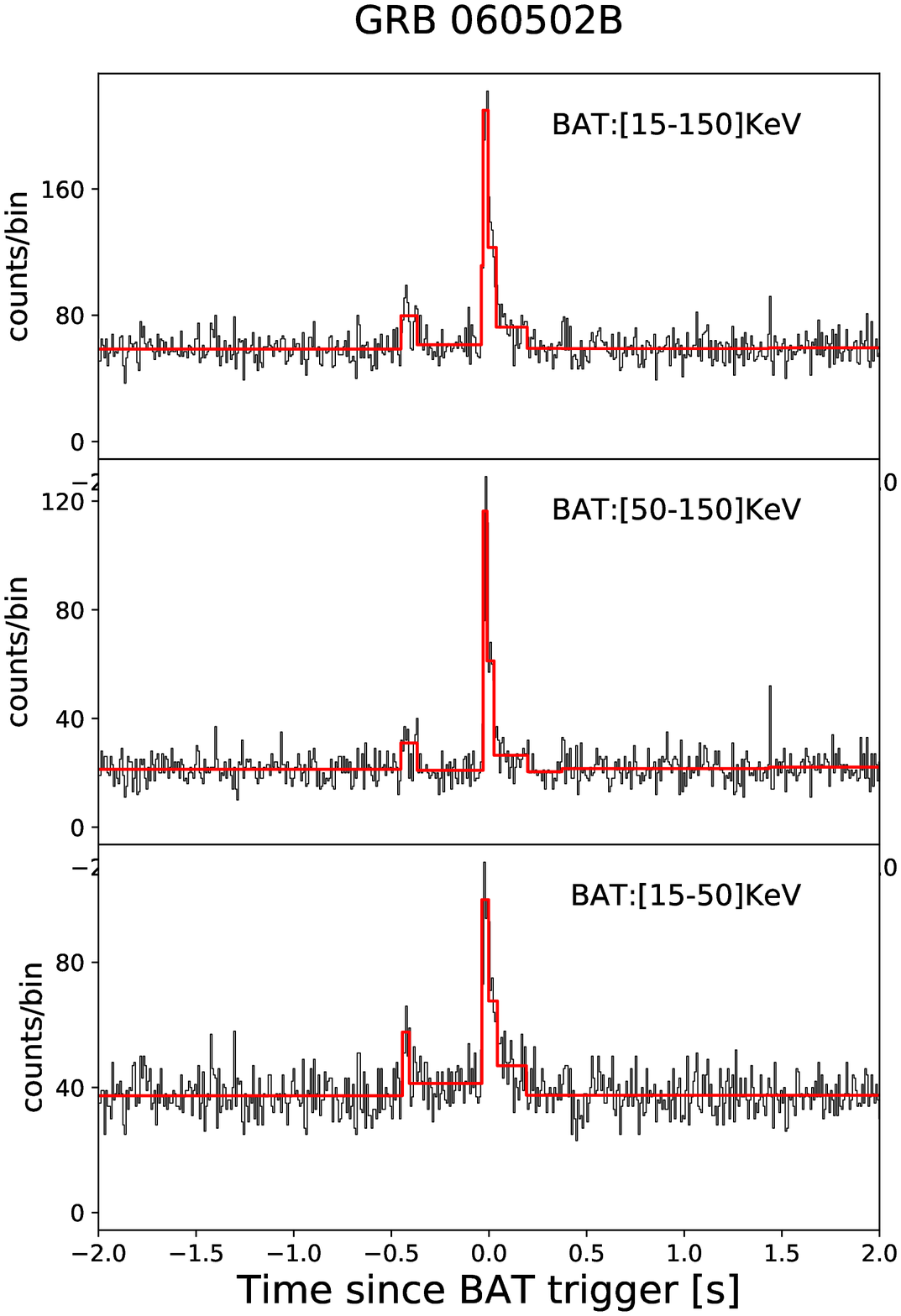}
\includegraphics[angle=0,width=0.5\textwidth]{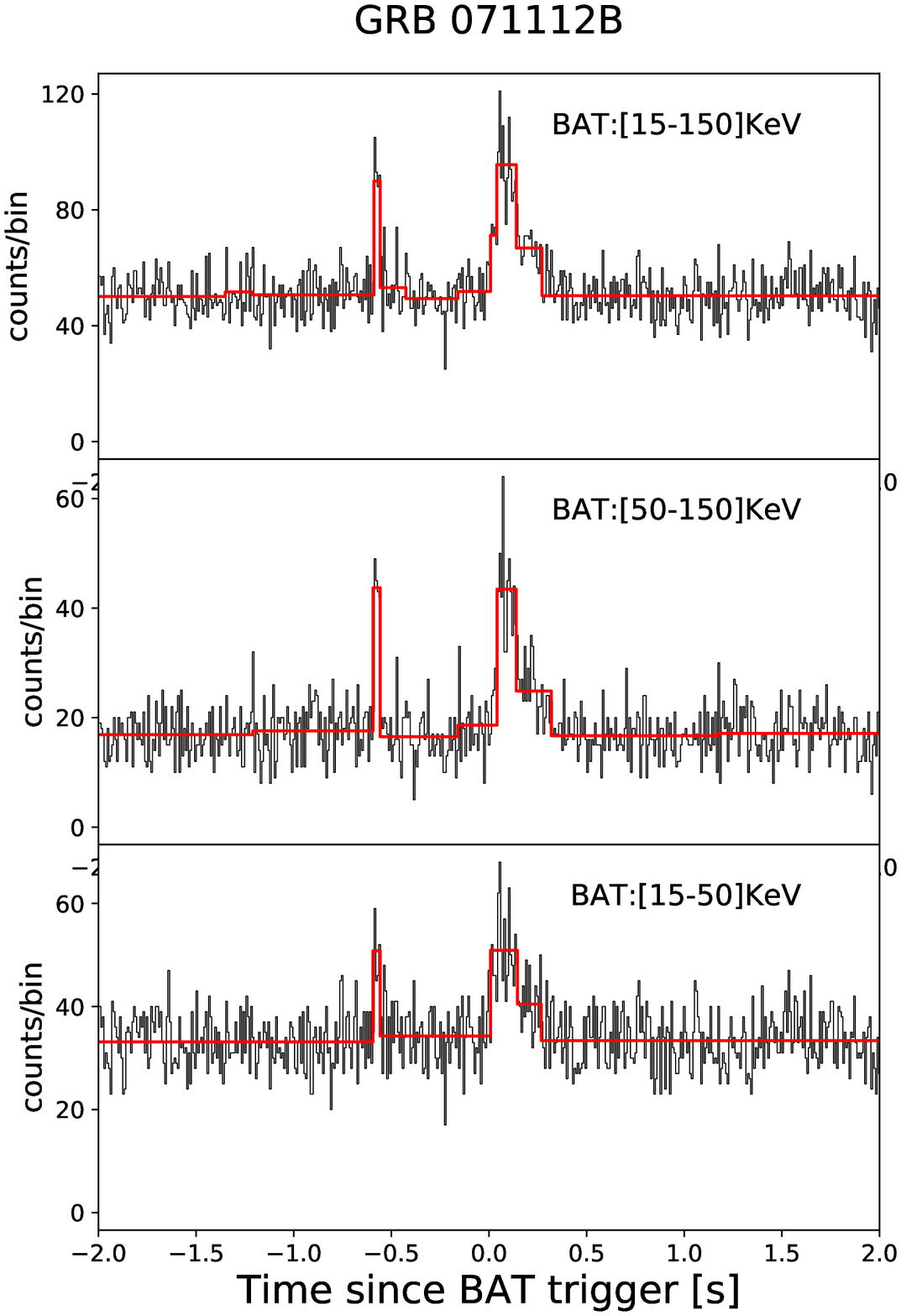}
\includegraphics[angle=0,width=0.5\textwidth]{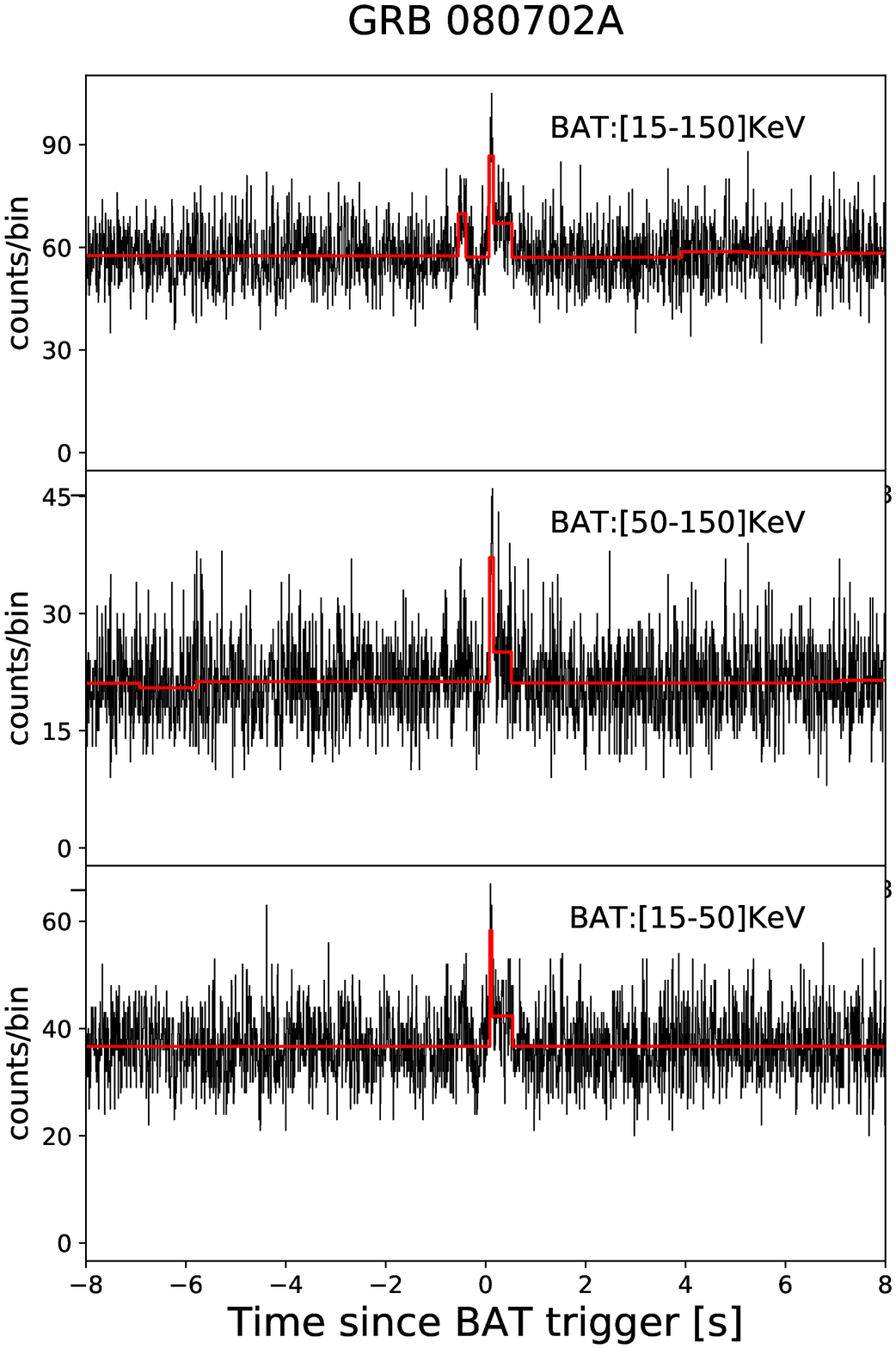}
\includegraphics[angle=0,width=0.5\textwidth]{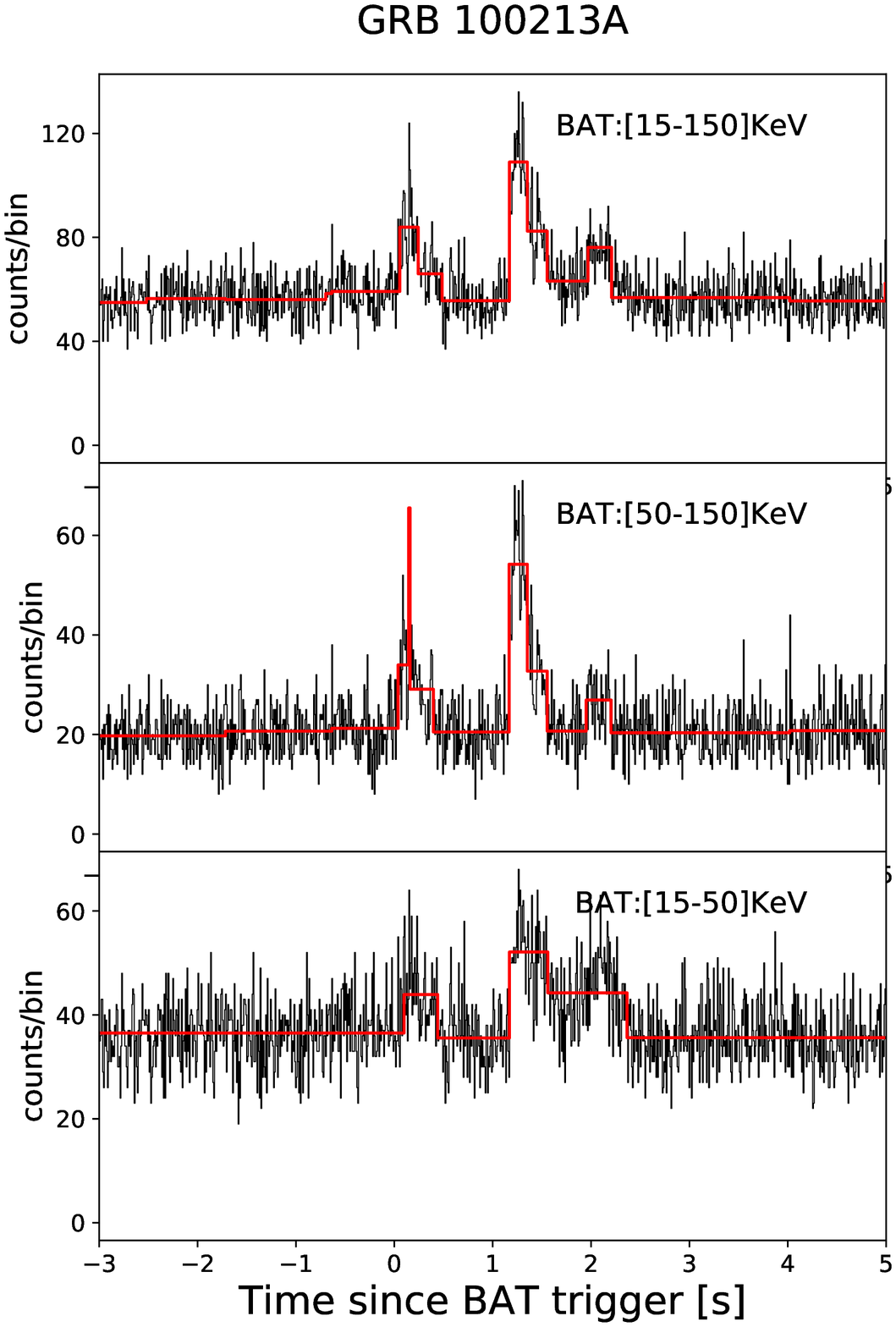}
\caption{The BAT light curves in {\em Swift} SGRBs with precursors are identified by the BBlocks analysis.}
\label{LC_Swift}
\end{figure}

After performing BBlocks, we selected those candidates with precursors should satisfy the following two criteria
according to \cite{tro10}: (a) the peak flux of a prior episode (precursor) is smaller than
the posterior episode (main burst); (b) the precursor flux returns to the background level before the start of the main burst.
Finally, we sampled 18 candidate SGRBs, in which 5 are collected from only {\em Swift} BAT instrument,
7 are observed by only {\em Fermi} satellite, and 6 are detected by both {\em Swift} and {\em Fermi}.
The sample has an overlap with \cite{tro10} and \cite{mp17}.
As shown in Figure \ref{LC_Swift} and the left panels of Figures \ref{LC_Spec(Swift+Fermi)} and \ref{LC_Spec(Fermi)}, the temporal profiles are plotted in units of counts/bin (bin=8 ms) for different energy bands, with the results by the BBlocks analysis illustrated by the red solid lines. For these lines, the preburst and postburst temporal backgrounds are fitted with the first block and the last block. The time intervals that blocks over the background level occupy are treated as the time intervals of signals, as listed in Columns 3 and 6 of Table \ref{tab:results}. Additionally, in each time interval, the amplitudes of blocks minus the background (i.e., the amplitude of the first block or the last block) are treated as the fluxes of signals. As illuminated in the following hardness ratio (HR) and average flux analysis of the precursor and the main burst, we can carry out some calculations according to the amplitudes of blocks in the time intervals of the precursor and the main burst, within different energy bands.

Due to the false positive rate $0.01$ in each precursor, one may expect 1 spurious detection for 18 precursors.
In order to examine whether these precursors are real or spurious detections, we also did a cross check with the observations of other spacecrafts. 7 precursors are confidently considered to be real in the whole sample. The precursors in GRBs 090510, 140209A, 160726A, and 180402A were observed simultaneously by {\em Swift} and {\em Fermi} by our BBlocks analysis. \cite{tro10} checked that the precursor of GRB 081024A was also observed by both {\em Swift} and {\em Fermi}. In addition, the precursors in GRBs 100717372 and 130310840 are detected by {\em Swift}, {\em Fermi}, {\em Messenger}, {\em Agile}, and {\em Fermi}, {\em KONUS}, {\em Suzaku}, {\em Messenger}, {\em HEND-Odyssey}, respectively \citep{mp17}. Others in \cite{tro10} and \cite{mp17} such as GRBs 091117 and 071030 cannot be downloaded from the {\em Swift} archive listing any more to re-analyze by the BBlocks technique. Several precursors such as in GRB 050724, in the time interval $-140.6\sim-139.5$ s of GRB 080702A, and in the time interval $-13.0\sim-12.6$ s of GRB 090510 reported by \cite{tro10}, cannot be identified by our BBlocks analysis for the following reasons. (1) Data type: \cite{tro10} used the RATE column of binned event data with a time bin of 0.128 ms, while we used the COUNTS column of TTE data. (2) Algorithm: a wavelet analysis with a Morlet mother function which has both time and frequency resolution, was adopted in \cite{tro10}; while the BBlocks is focused on detecting the time-domain shape of signals with no reference to any
frequency limitations or behavior. Those precursors identified by a wavelet analysis rather than the BBlocks are likely attributed to the frequency (periodicity) resolution. Nevertheless, we can also see that the precursors in the time interval -0.56$\sim$-0.25 s of GRB 080702A and in GRBs 060502B and 071112B in {\em Swift} before 2010 were not found by \cite{tro10}.

\subsection{Image Analysis}
\label{subsec:imagine}

As did in \cite{tro10}, we also used {\em batcelldetect} operating on background-subtracted sky image and partial coding map to perform a source detection for the time interval of precursor at its corresponding GRB position, for all candidate SGRBs observed by {\em Swift}. The handling process refers to BAT analysis threads\footnote{https://swift.gsfc.nasa.gov/analysis/threads/bat\_threads.html}. The significance levels of the precursors in the image domain are listed in the last column of Table \ref{tab:results}. For a blind source detection considered to be real, it should have a significance threshold of at least $6.5\sigma$ across the entire BAT detector area. As we already knew the GRB position, the blind source detection in the image domain is not necessary. Instead, detecting a source at a known position can reduce the
number of trials by a factor of $3\times10^4$, so the values of significance in the last column would be more restrictive than the same values in blind search. Moreover, the precursors such as in GRBs 090510 and 180402A are confidently deemed to be real since they are detected by both {\em Swift} and {\em Fermi}, though their significance levels in the image domain seem low.

\section{Results}
\label{sec:results}
\subsection{Temporal Properties}
\label{subsec:temporal properties}

The temporal light curves of 18 candidates with precursor emission are illustrated in Figure \ref{LC_Swift}, and the left panels of Figures \ref{LC_Spec(Swift+Fermi)} and \ref{LC_Spec(Fermi)}, where the red solid block patterns over background show signals.
Among them, one faint precursor signal in GRB 130310840 is magnified to show clearly with the inset.
Through the BBlocks analysis for the light curves, we defined the time interval that the blocks over background occupy for each
episode as its duration (the precursor $T_{\rm pre}$ and the main burst $T_{\rm main}$),
and the time interval from the end of the precursor to the beginning of
the main burst as the quiescent time $T_{\rm quie}$. Since the block light curves
are separated into different energy bands, we mainly chose the whole band 15-150 keV
for BAT light curves and 8-1000 keV for GBM light curves to determine the durations.
If the precursor signature does not emerge in the whole energy band light curves such as GRBs 100827455 and 130310840, however,
we would choose the other energy band light curve with the strongest precursor emission to make an analysis.
The results of the time intervals are reported in Table \ref{tab:results}. We further made distributions for these durations by using a Gaussian function. As shown in the {\em left panel} of Figure \ref{Time}, the precursors distribute as
log$(T_{\rm pre}/{\rm s})=-0.91\pm0.99$, which is obviously shorter than that of the main bursts
log$(T_{\rm main}/{\rm s})=-0.29\pm0.65$. The quiescent time distribution log$(T_{\rm quie}/{\rm s})=-0.32\pm0.21$,
with a relatively small scatter, is comparable to the main burst.
For the whole sample, the precursor durations, main burst durations, and quiescent times span the ranges
0.01-0.6 s, 0.08-2.39 s, and 0.13-3.65 s, respectively. More specifically,
the precursors in a major fraction of (16/18) cases are shorter than the corresponding main bursts.
In contrast, the quiescent times in more than half of (11/18) cases are longer than their corresponding main bursts.
Moreover, we also did correlation analysis for the precursor duration,
main burst duration, and quiescent time, but found no any significant correlations among them, as seen in the {\em right panel} of Figure \ref{Time}.
Furthermore, it is worth noting that there is no quiescent time $> 4$ second in our sample by the BBlocks analysis,
compared with the quiescent time up to 100 second in \cite{tro10}. This could lead to the differences such as in data type and algorithm as aforementioned in the end of \S \ref{subsec:sample selection}. Otherwise, \cite{mp17} did not yet found those cases that have a quiescent time $>5$ second in the {\em SPI-ACS/INTEGRAL} experiment via a wavelet analysis. Therefore, it is unclear why our result differs from that of \cite{tro10}.

Although we selected the candidate SGRBs with precursors in terms of the peak flux difference between
the less-intensive episode and the proper episode, we also calculated
the average flux for each episode. The average flux can be estimated as the average amplitude of blocks over background in the time interval of episode, from the BBlock analysis within the energy band 15-150 keV for SGRBs observed by only {\em Swift} and by both {\em Swift} and {\em Fermi}, and within the energy band 8-1000 keV for those observed by only {\em Fermi}. The standard deviation of block amplitudes is regarded as the average flux error in each episode. Illustrated in the {\em right panel} of Figure \ref{hardness}, the average flux for each precursor is also generally smaller than that of its main burst.
Additionally, the average flux of precursors tends to increase as their main bursts brighten,
with log$F_{\rm main}=(0.47\pm0.19)\times$log$F_{\rm pre}+(0.93\pm0.20)$
(Pearson correlation coefficient $r=0.53$ and chance probability $p<0.03$),
suggesting a weak positive relation in average flux between the precursors and the main bursts.

\subsection{Spectral Properties}
\label{subsec:spectral properties}

The HR is defined as the ratio of count rate in 50-150 keV over 15-50 keV for SGRBs observed by {\em Swift} and by both {\em Swift} and {\em Fermi}. For SGRBs observed by only {\em Fermi}, the HR is defined as the count rate in 50-1000 keV over 8-50 keV. This is an applicable quantity to illuminate the spectral feature especially for faint emission and for narrow {\em Swift} BAT spectra which have only a sensitive energy range 15-150 keV. We proceeded a HR calculation for the precursors and main bursts identified by our BBlocks analysis. Apparently, the HR can also be estimated from the average flux ratio (the average amplitude ratio of blocks) in different energy bands shown in Figures \ref{LC_Swift}, \ref{LC_Spec(Swift+Fermi)}, and \ref{LC_Spec(Fermi)}, after the background subtraction from the first block and the last block. The HR error is treated as the standard deviations of amplitudes of blocks in two different energy bands. As displayed in the {\em left panel} of Figure \ref{hardness}, the HR Gauss distributions are log$HR_{\rm pre}=-0.03\pm0.49$
for the precursors and log$HR_{\rm main}=0.06\pm0.34$ for the main bursts, indicating that the main bursts are slightly harder than the precursors. However, we can also see that a half of precursors occupy the region $HR_{\rm pre}>HR_{\rm main}$, no matter for candidates observed by only {\em Swift} and by both {\em Swift} and {\em Fermi} or for those observed by only {\em Fermi}. Furthermore, there is no any relation in HR between the precursors and the main bursts to be found.
These results are basically in agreement with the HR results in \cite{tro10}.

As the {\em Swift} BAT has a narrow energy band, its spectra usually can be well fit
with only single power-law (PL) model. In contrast,
the {\em Fermi} GBM and LAT cover a wide energy range in gamma-ray bands and their GRB spectra usually appear
one or two breaks and can be well fit with a cutoff power-law model (CPL) (or adding a PL model) \citep{lu17,lan18},
read
\begin{equation}
N_{\rm CPL}(E)=N_{\rm 0,CPL}E^{-\Gamma_{\rm CPL}}{\rm exp}\left (
\frac{E}{E_{\rm c}}\right),
\end{equation}
where $\Gamma_{\rm CPL}$, $N_{\rm 0,CPL}$, and $E_{\rm c}$ are the photon spectral index,
the normalization parameter in units of photons/keV/cm$^2$/s at 1 keV, and the cutoff energy in keV, respectively.
Thus we did a spectral fitting only for those candidates observed by {\em Fermi} and by both {\em Fermi} and {\em Swift}.
Additionally, a blackbody (BB) spectrum function is also used for trial fitting to the spectra in order to search for the thermal component,
which is expressed by
\begin{equation}
N_{\rm BB}(E)=N_{\rm 0,BB}\left(\frac{8.0525E^2~dE}{(kT)^4[{\rm exp}(E/kT)-1]}\right),
\end{equation}
where $kT$ is the temperature in keV.
These two models are built in {\em Xspec}. We also used the tools {\em gtburst} to extract
the GBM and LAT-LLE spectrum files, and {\em Xspec} to fit the GBM and LAT-LLE spectra,
combining the BAT spectra. A PGSTAT statistics method is used for GBM and LAT-LLE spectrum fitting \citep{cash79},
while the default chi-squared (CHI) is invoked for BAT spectrum fitting. The reduced $\chi^2$ is given to
estimate the goodness of fitting results. The comparison for the fitting goodness between CPL and BB model,
is made by the Bayesian information criterion (BIC), for which evidence against the model
with the higher $\Delta$BIC can be formulated as: a) $0\sim2$, not worth more than a bare mention (NM);
b) $2\sim6$, positive (P); c) $6\sim10$, strong (S); d) $>10$, very strong (VS).
The detail please refer to \cite{lv17a}.

The time-averaged spectra are extracted and analyzed for the precursors and main bursts.
As shown in the right columns of Figures \ref{LC_Spec(Swift+Fermi)} and \ref{LC_Spec(Fermi)},
the best fitting results are plotted with a top label indicating the time interval and spectral model.
For the SGRBs observed by only {\em Fermi} satellite, we used the data of two NaI and one
BGO detectors to carry out a spectral fitting. For those observed by both {\em Swift} and {\em Fermi},
we used the data of BAT, one NaI, and one BGO (and LAT-LLE, if any).
Note that GRB 090510 presents a very high energy LAT-LLE emission in main event,
its spectral fitting requires a CPL with an extra PL model.

From the BIC analysis, the better fitting model CPL accounts for the spectra
of all main bursts of those candidates observed by only {\em Fermi} and by both {\em Swift} and {\em Fermi}.
For the precursors, a significant portion are suitable for fitting with also a CPL instead of a BB model, as covered in Table \ref{tab:spectral fitting}.
To do a spectral comparison,
we employed the cutoff energy $E_{\rm c}$ to analyze the spectral feature of the precursors and main bursts.
As illustrated in the {\em left panel} of Figure \ref{spec_Ec}, the cutoff energy Gauss distributions are log$(E_{\rm c,pre}/{\rm keV})=2.54\pm0.44$
for the precursors and log$(E_{\rm c,main}/{\rm keV})=2.86\pm0.71$ for the main bursts.
On the whole, these distributions show that the main bursts are slightly spectrally harder with respect to the precursors. But for about a half of cases, the precursors are spectrally harder than their main bursts, as seen from the cutoff energy of the main bursts as a function of that of the corresponding precursors.
In addition, there is no any significant relation for the cutoff energy $E_{\rm c}$ between the precursors and the main bursts.
These results are all consistent with those of HR comparison.
Furthermore, we also investigated the potential correlation between the cutoff energy $E_c$ and the average flux $F$ for the precursors and main bursts but found no any correlation between them, revealed in the {\em right panel} of Figure \ref{spec_Ec}.

\section{Discussion}
\label{sec:discussion}
\subsection{Precursor: a Cocoon Emission?}
\label{subsec:cocoon}

The spectral properties of precursors in LGRBs have been investigated by several
systematic searches with {\em BATSE} \citep{kos95,laz05}, {\em Swift} \citep{bur08,hu14}, and {\em Fermi} observations \citep{lan18}.
Some authors showed that the precursors are commonly non-thermal and also
spectrally softer than their main bursts \citep{laz05}, while some other authors suggested that
the spectra of precursors are neither systematically harder nor softer than the main ones \citep{bur08,hu14}.
For the sake of a wide energy range of {\em Fermi} data, we can also use the cutoff energy $E_{\rm c}$ to
assess the spectral features, besides the HR. The cutoff energy distributions of precursors and main bursts as well as their HR distributions indicate the slightly softer precursors with respect to the main ones. But there are a half of candidates that their precursors are spectrally harder than their corresponding main bursts, from both the comparisons of cutoff energy and HR between precursors and main bursts. Moreover, our results showed that
a large portion of precursors have a non-thermal spectrum, analogous to their main bursts.

Theoretically, precursors in both LGRBs and SGRBs are difficult to interpret
within the standard fireball scenario because of the requirement that at least the main burst duration
is longer than the quiescent time \citep{laz05,tro10}, in striking contrast
to our results in Figure \ref{Time} and Table \ref{tab:results}. Other explanations for precursors in LGRBs
are associated with the collapsar scenario \citep{ram02,lb05}.
Among them, even the seemingly most plausible cocoon emission model might not be suitable for the precursors in our SGRB sample.
On one hand, the burst SGRB 170817A produced from the merger of NS-NS is very possibly involved
to the off-axis structured jet rather than the jet-cocoon geometry due to the afterglow observations \citep{tro18},
the polarization measurements and/or imaging about its source \citep{moo18,gill18},
and especially the recent high resolution measurements of the source size and position \citep{ghi19}.
Hence, the emergence of the cocoon that produces the precursor emission, stemmed from the merger of NS-NS,
maybe encounter a challenge. On the other hand, the cocoon emission is widely believed
to be associated with a nearly thermal spectrum \citep{laz17,np17,de18},
cannot account for a large fraction of precursors with a non-thermal spectral feature in our sample.

\subsection{Precursor: Magnetospheric Interaction or Tidal Crust Cracking?}
\label{subsec:magnetospheric}

If the precursor emission in SGRBs comes from the interactions of inspiral binary compact
stars (NS-NS or NS-BH) at the final stage prior to coalescence, such as the magnetospheric
interaction \citep{vie96,han01,mcw11,pal13,wang18}
or the tidal crust cracking \citep{tro10,tsa12}, the precursor might provide
a probe for the nature of progenitors of SGRBs. The NS in binary compact stars is usually magnetized strongly,
a strong voltage and accelerated charged particles would be induced when
the company NS or BH crosses its dipole magnetic field \citep[see a review][]{fer16}.
The accelerated particles would produce observable electromagnetic emission in X-ray/$\gamma$-ray bands.
Under the magnetosphere inflation condition in which the toroidal magnetic field induced by the current becomes comparable to
that of the original poloidal dipole field, we can estimate the maximum energy release
as the orbit decays to separation $a$ with equation \citep{con99,lai12,fer16},
\begin{equation}
E_{\rm tot}\approx~6\times10^{42}~{\rm erg}~\left(\frac{B_{\rm d}}{10^{13} {\rm G}}\right)^2~\left(\frac{a}{2R_{\rm ns}}\right)^{-3}
\end{equation}
where $B_{\rm d}$ and $R_{\rm ns}$ are the surface equatorial dipole field of a magnetized NS and its radius, respectively.
In our sample, only GRBs 060502B and 090510 have a redshift measure.
Adopting the precursor duration 0.04 s and its average flux integrated over the whole energy range in the CPL fitting model $(2.5\pm0.3)\times10^{-6}~{\rm erg~cm^{-2}~s^{-1}}$,
we can calculate its fluence and thus the total radiation energy $E_{\rm tot}=(6.8\pm0.9)\times10^{50}$ erg for GRB 090510.
As for GRB 060502B, its precursor average flux 0.15 ${\rm counts~s^{-1}~cm^{-2}}$
in the 15 - 150 keV energy band and duration $\sim$0.1 s from the BBlocks results,
we can calculate its fluence with an assumed Band function \citep{band93} and $K$-correction.
We used a PL model to fit the spectrum of precursor in the energy band 15-150 keV and obtained the low-energy photon spectral index $\alpha=-\Gamma_{\rm PL}=-1.79$
and the spectral peak energy $E_{\rm p}=70$ keV of $\nu f_{\nu}$ spectrum \citep[see an empirical relation ${\rm log}E_{\rm p}=(2.76\pm0.07)-(3.61\pm0.26)~{\rm log}\Gamma_{\rm PL}$ in][]{zhang07}.
Moreover, we assumed a typical high-energy photon spectral index $\beta=-2.3$.
We can finally estimate its total precursor energy $E_{\rm tot}=2.0\times10^{42}$ erg.
Therefore, assuming the separation up to the point of merger $a=2R_{\rm ns}$,
we can constrain the magnetic field strength of the magnetized NS in progenitor,
with an extremely strong magnetar-like magnetic field $B_{\rm d}\geq 1.1\times10^{17}$ G for GRB 090510
and a normally strong radio pulsar magnetic field $B_{\rm d}\geq 5.8\times10^{12}$ G for GRB 060502B.
Further, \cite{fer16} also estimated the precursor characteristic duration is approximately 1-10 ms based on the maximum released energy and the Poynting luminosity of the rotating magnetized NS in compact binary, which is basically consistent with the precursor duration distribution in our sample. This supports the magnetospheric interaction origin of precursor in SGRBs.

If the precursor is associated with the tidal crust cracking without resonant excitation of modes,
ascribing to the tidal distortion, the duration of tidal grinding prior to the merger would be \citep{tro10}
\begin{equation}
\tau_{\rm tg}\approx~3720~{\rm sec}~\epsilon_{-6}^{-4/3}~\frac{2q^{1/3}}{1+q}\left(\frac{m_{\rm ns}}{1.4~M_{\odot}}\right)^{-3}~\left(\frac{R_{\rm ns}}{10~ {\rm km}}\right)^{4}
\end{equation}
where $\epsilon$, $m_{\rm ns}$, $R_{\rm ns}$, and $q=m_2/m_1$ are the NS's ellipticity,
mass, radius, and the binary mass ratio, respectively.
Assuming the precursor arises simultaneously when the first crust cracking occurs,
the duration $\tau_{\rm tg}$ can be equal to the sum of the precursor duration and the quiescent time,
written by (ignoring the cosmological redshift here)
\begin{equation}
\tau_{\rm tg}=T_{\rm pre}+T_{\rm quie}
\end{equation}
Supposing the typical values $\epsilon\sim10^{-6}$ \citep{owen05}, $m_{\rm ns}\sim1.4~M_{\odot}$,
and $R_{\rm ns}\sim10~{\rm km}$, one would obtain the binary mass ratio $q$ as large as $10^{5}$
from the duration distribution $\tau_{\rm tg}\sim0.25 - 4.23$ s. Obviously, this is unreasonable.
Hence, at least the precursors in our sample are unlikely to arise from the tidal crust cracking
without resonant excitation of modes. Based on \cite{tsa12} instead,
the resonant shattering in NS crust might explain our results.
From the range of the duration $\tau_{\rm tg}$, we can see that NS
disfavors the crustal equations of state such as GS and RS \citep{fri86,ste09}
for chirp masses of 1.0-4.5$M_{\odot}$ according to Fig. 2 in \cite{tsa12}.

\section{Conclusions}
\label{sec:conclusions}

We have performed a comprehensive search for the candidate SGRBs with precursors from
nearly 120 {\em Swift} and 540 {\em Fermi} SGRBs through the BBlocks technique.
18 SGRBs observed by {\em Swift} and/or {\em Fermi} satellites that
satisfy the precursor criteria have been selected. By analyzing their temporal profiles,
we determined the durations of their precursors, main bursts, and quiescent times,
studied their distributions, and sought the possible relations among them. We used HR to analyzed the spectral feature of the precursors and their main bursts. Moreover, we also adopted CPL,
and BB models to fit the spectra of precursors and main bursts, and investigated their spectral properties. Via the BIC method,
we further compared the fitting goodness between CPL and BB modelling the
spectra of precursors and main bursts. Finally, we explored the origin
of precursors in SGRBs and their possible probes for the progenitors.
The meaningful results we acquired are as follows:
\begin{itemize}
\item The precursor durations have a distribution log$(T_{\rm pre}/{\rm s})=-0.91\pm0.99$,
being shorter than the main burst log$(T_{\rm main}/{\rm s})=-0.29\pm0.65$ and the
quiescent time log$(T_{\rm quie}/{\rm s})=-0.32\pm0.21$. Besides, the quiescent times are comparable to the main burst durations. For the whole sample, the precursor durations, main burst durations, and quiescent times span the ranges
0.01-0.6 s, 0.08-2.39 s, and 0.13-3.65 s.
More specifically, for a major part of cases, their precursor durations are
shorter than their main bursts.
Moreover, there seem to be no any relations among the quiescent
time, the precursor duration, and the main burst duration.
Additionally, the average flux of precursors tends to increase as the main bursts brighten,
with log$F_{\rm main}=(0.47\pm0.19)\times$log$F_{\rm pre}+(0.93\pm0.20)$,
suggesting a weak positive relation in the average flux between the precursors and their main bursts.
\item The spectral feature HR distributions are log$HR_{\rm pre}=-0.03\pm0.49$
for the precursors and log$HR_{\rm main}=0.06\pm0.34$ for the main bursts, indicating the main bursts are slightly harder than the precursors. Nevertheless, we can also see that a half of precursors occupy the region $HR_{\rm pre}>HR_{\rm main}$. In addition, the spectral fitting results by CPL and BB models for those candidates observed by {\em Fermi} indicate that a large portion of precursors and all main bursts favor a non-thermal spectrum.
Another spectral quantity cutoff energy $E_c$ in CPL fitting distribute as log$(E_{\rm c,pre}/{\rm keV})=2.54\pm0.44$
for the precursors and log$(E_{\rm c,main}/{\rm keV})=2.86\pm0.71$ for the main bursts. This also shows that the main bursts are slightly spectrally harder than the precursors. But for about a half of these candidates, their precursors are
spectrally harder than their main bursts. It is consistent with the result of HR comparison.
\item If the precursor emerges in the post-merger phase,
it might originate from the cocoon emission rather than a fireball
becoming optically thin, as the quiescent time is longer
than the corresponding main burst duration for more than a half of candidates. Whereas the cocoon emission is usually
believed to be related to a thermal spectrum, a significant portion of precursors favoring a non-thermal spectrum in our sample would challenge this interpretation.
On the other hand, the emergence of the cocoon resulting from the merger
of binary NSs may also encounter a challenge since the
jet-cocoon origin of SGRB 170817A could be excluded from the recent observations \citep{ghi19}.
\item If the precursor occurs prior to the merger, it is likely
associated with the magnetospheric interaction or the tidal crust
cracking with resonant excitation of modes. In the case of magnetospheric interaction,
the precursor can be used to constrain the surface equatorial dipole field of magnetized NS in progenitor.
For GRBs 090510 and 060502B with a redshift measure, their magnetized NSs
in progenitors reach to a magnetar-like magnetic field $1.1\times10^{17}$ G
and a radio pulsar magnetic field $5.8\times10^{12}$ G, respectively.
Under the condition that the precursor is related to the tidal crust cracking with resonant excitation of modes,
we can see that NS disfavors the crustal equations of state such as GS and RS
from the duration of tidal grinding prior to the merger $\tau_{\rm tg}$ with only several seconds \citep{tsa12}.

\end{itemize}

Under the condition of lacking the gravitational wave observations,
the identification of progenitors (NS-NS or NS-BH) in SGRBs is usually
difficult to be done through the electromagnetic observations post merger phase,
although some electromagnetic signatures may provide hints about the types of progenitors.
One is the X-ray plateau followed by a sharp decay phase in some afterglows of SGRBs,
generally believed to be due to a long-lasting magnetar
originating from the merger of NS-NS \citep{tro07,row13,lv15,lv17b,xue19},
following the early suggestion \citep{Dai98a,Dai98b,zhang01,Dai04}.
The other is the giant X-ray flare in afterglows of a few SGRBs, which is possibly
caused by the merger of NS-NS \citep{Dai06} or NS-BH \citep{mu18}. The precursor in SGRBs,
if it is indeed associated with the progenitor and seen to be in difference
between NS-NS and NS-BH mergers, could be a new pre-merger probe for
the identification of progenitors, with the sample extension of SGRBs with
precursor benefiting from more future observations by {\em Fermi} and
{\em SVOM} satellites \citep{pau11}. Furthermore, \cite{han01} and \cite{wang16}
pointed out the precursors in SGRBs are likely associated with the radio emissions
like fast radio bursts (FRBs), so we collected FRBs from the catalog\footnote{http://www.frbcat.org}
\citep{pet16} updated to date and found no any position consistency
between these FRBs and precursors in these SGRBs.

\acknowledgments

We acknowledge the use of the public data from the {\em Fermi} and {\em Swift} data archive. This work is supported by the National Key Research and Development Program of China (grant No. 2017YFA0402600) and the National Natural Science Foundation of China (grant No. 11573014 and 11833003).



\begin{deluxetable}{ccccccccccccc}
\tablewidth{0pt}
\tabletypesize{\tiny}
\tablecaption{Results of SGRBs with Candidate Precursors}
\tablehead{
\colhead{GRB$^{a}$} &
\colhead{z} &
\colhead{Presursor$^{b}$} &
\colhead{$F_{\rm pre}$$^{c}$} &
\colhead{$HR_{\rm pre}$$^{d}$} &
\colhead{Main Burst$^{b}$} &
\colhead{$F_{\rm main}$$^{c}$} &
\colhead{$HR_{\rm main}$$^{d}$} &
\colhead{$T_{\rm quie}$$^{e}$} &
\colhead{Significance$^{f}$} \\
\colhead{} &
\colhead{} &
\colhead{(s)} &
\colhead{(counts/bin)} &
\colhead{} &
\colhead{(s)} &
\colhead{(counts/bin)} &
\colhead{} &
\colhead{(s)} &
\colhead{($\sigma$)}
}
\startdata
\object{\em Swift} &&&&&&&&& \\
\hline
\object{060502B} &0.287&-0.45$\sim$-0.36&20&0.38$\pm$0.38&-0.04$\sim$0.20&59$\pm$51&1.20$\pm$2.19&0.32&6.1 \\
\object{071112B} & &-0.60$\sim$-0.59&14&2.63&0.00$\sim$0.27&24$\pm$13&1.28$\pm$1.40&0.59&1.9 \\
\object{080702A} & &-0.56$\sim$-0.25&8&0.80&-0.12$\sim$0.52&15$\pm$12&0.62$\pm$1.07&0.13&3.3 \\
\object{100213A} & &0.05$\sim$0.49&42$\pm$39&0.49$\pm$1.09&1.17$\sim$2.21&26$\pm$16&0.95$\pm$1.17&0.68&11.1 \\
\object{100702A} & &-0.27$\sim$-0.23&27&0.31&0.00$\sim$0.21&57$\pm$25&0.67$\pm$0.64&0.23&4.8 \\
\hline
\object{\em Swift+Fermi} &&&&&&&&& \\
\hline
\object{081024A} &&-1.65$\sim$-1.59&36$\pm$6&1.37$\pm$0.19&-0.68$\sim$0.26&22$\pm$22&0.98$\pm$1.80&0.91&6.7 \\
\object{090510} &0.903&-0.55$\sim$-0.51&26$\pm$23&1.34$\pm$1.67&-0.11$\sim$0.45&96$\pm$72&1.49$\pm$2.54&0.40&4.9 \\
\object{140209A} &&0.12$\sim$0.57&16&0.43&1.63$\sim$4.02&147$\pm$115&1.28$\pm$2.15&1.06&13.9 \\
\object{160408A} &&-0.94$\sim$-0.87&0.8&0.31&-0.33$\sim$0.41&12$\pm$8&0.59$\pm$0.71&0.55&5.2 \\
\object{160726A} &&0.29$\sim$0.37&25$\pm$32&0.60$\pm$1.55&0.76$\sim$1.08&63$\pm$44&1.00$\pm$1.71&0.39&10.2 \\
\object{180402A} &&-0.20$\sim$-0.17&23&3.61&0.05$\sim$0.29&44$\pm$29&2.23$\pm$1.45&0.22&2.5 \\
\hline
\object{\em Fermi} &&&& &&&&& \\
\hline
\object{081216531} &&-0.14$\sim$-0.01&6.6&0.96&0.50$\sim$1.00&13$\pm$10&2.45$\pm$3.34&0.51&... \\
\object{100717372} &&-0.14$\sim$0.01&3.4&5.15&3.33$\sim$4.56&3.2&3.55$\pm2.68$&3.32&... \\
\object{100827455} &&-0.06$\sim$-0.01&6.5&4.85&0.39$\sim$0.51&15&3.95&0.40&... \\
\object{130310840} &&-0.18$\sim$0.40&1.2&52.59&4.05$\sim$6.20&56$\pm$65&1.10$\pm$2.61&3.65&... \\
\object{160818198} &&-0.42$\sim$0.18&4.1&0.61&2.58$\sim$4.05&6.5&0.64$\pm$0.76&2.40&... \\
\object{170709334} &&-0.15$\sim$0.27&3.1&3.44&0.67$\sim$0.75&13&4.02&0.40&... \\
\object{170726794} &&-0.18$\sim$0.10&3.6&2.76&1.63$\sim$1.88&6.1&1.87&1.53&... \\
\hline
\enddata
\tablenotetext{a}{The SGRBs observed by both {\em Swift} and {\em Fermi}: GRB 081024A-bn081024245,
GRB 090510-bn090510016, GRB 140209A-bn140209313, GRB 160408A-bn160408268, GRB 160726A-bn160726065, GRB 180402A-bn180402406.}
\tablenotetext{b}{The time intervals of the precursors and main bursts. The intervals of SGRBs observed by only {\em Swift} and by both {\em Swift} and {\em Fermi} are all since BAT trigger, while those observed by only {\em Fermi} are since GBM trigger.}
\tablenotetext{c}{Average flux of the precursors and main bursts in units of counts/bin (bin = 8 ms) in the energy band 15-150 keV for SGRBs observed by only {\em Swift} and by both {\em Swift} and {\em Fermi}, and the energy band 8-1000 keV for SGRBs observed by only {\em Fermi}. The flux standard deviation is treated as the flux error. Some flux errors are not listed due to their extremely small values.}
\tablenotetext{d}{Hardness ratio (HR) is defined as the count rate in the 50-150 keV band over the count rate in the 15-50 keV band for SGRBs observed by only {\em Swift} and by both {\em Swift} and {\em Fermi}, while for those observed by only {\em Fermi}, their HRs are defined as the count rate in the 50-1000 keV band over the count rate in the 8-50 keV band. The HR error is treated as the count rate standard deviation transfer. Some errors are not listed due to their extremely small values.}
\tablenotetext{e}{The quiescent times from the end of the precursor to the beginning of the main burst.}
\tablenotetext{f}{Image significance of the candidate precursors via performing source detections for SGRBs observed by only {\em Swift} and by both {\em Swift} and {\em Fermi}.}
\label{tab:results}
\end{deluxetable}

\clearpage
\begin{deluxetable}{cccccccccccccccc}
\rotate
\tablewidth{0pt}
\tabletypesize{\tiny}
\tablecaption{Spectral Fittings with PL, CPL, and BB
 Models for Precursor and Main Burst}
\tablenum{2}
\tablehead{
\colhead{GRB} &
\colhead{ } &
\colhead{ } &
\colhead{ } &
\colhead{\rm precursor } &
\colhead{ } &
\colhead{ } &
\colhead{ } &
\colhead{ } &
\colhead{ } &
\colhead{ } &
\colhead{\rm main burst } &
\colhead{ } &
\colhead{ } &
\colhead{ }
}

\startdata
\object{\em Swift+Fermi} &$\Gamma_{\rm CPL}$&$E_{\rm c}$ (keV)&${\chi^2/\rm dof}$&$kT$ (keV)&${\chi^2/\rm dof}$&$\Delta \rm BIC$ & BIC-selected model$^{a}$&$\Gamma_{\rm CPL}$ &$E_{\rm c}$ (keV)&${\chi^2/\rm dof}$&$kT$ (keV)&${\chi^2/\rm dof}$&$\Delta \rm BIC$&BIC-selected model \\
\hline
\object{081024A} &...&...&...&...&...&...&...&$1.16\pm0.14$
&$2390\pm2440$&339/294&$20\pm2$&378/295&-33&CPL(VS) \\
\object{090510} &$0.39\pm0.45$ &$249\pm167$ &232/342 &$55\pm8$ &239/343 &-1 &CPL(NM) &$0.75\pm0.06$
&$4316\pm342$&296/369&$537\pm19$&509/370&-207&CPL+PL(VS) \\
\object{140209A} &$1.54\pm0.16$ &$1479\pm1456$ &300/296 &$14\pm1$ &337/297 &-31 &CPL(VS) &$0.97\pm0.03$  &$214\pm14$&434/296&$23\pm1$&2679/297&-2239&CPL(VS) \\
\object{160408A} &...&...&...&...&...&...&...&$0.66\pm0.13$
&$495\pm179$&327/298&$69\pm4$&405/299&-27&CPL(VS) \\
\object{160726A} &$0.88\pm0.21$ &$342\pm188$ &276/297 &$29\pm3$ &309/298 &-27 &CPL(VS) &$1.05\pm0.08$
&$585\pm378$&269/297&$36\pm2$&461/298&-186&CPL(VS) \\
\object{180402A} &$-0.05\pm0.41$ &$978\pm683$ &247/296 &$112\pm30$ &248/297 &-17 &CPL(VS) &$0.39\pm0.09$
&$778\pm132$&299/296&$126\pm6$&401/297&-90&CPL(VS) \\
\hline
\object{\em Fermi} &&&&&&&&&&&&&& \\
\hline
\object{081216531} &$0.12\pm0.57$ &$59\pm32$ &339/357 &$22.8\pm2.5$ &342/358 &3 &BB(P) &$0.64\pm0.06$
&$779\pm118$&407/357&$100\pm3$&664/358&-251&CPL(VS) \\
\object{100717372} &$0.06\pm0.45$ &$308\pm160$ &361/356 &$100\pm12$ &367/357 &0 &CPL(NM) &$1.06\pm0.08$
&$3001\pm1112$&360/356&$100\pm8$&448/357&-82&CPL(VS) \\
\object{100827455} &$0.65\pm0.39$ &$331\pm318$ &329/358 &$47.1\pm7.4$ &339/359 &-4 &CPL(P) &$0.37\pm0.13$
&$464\pm106$&331/358&$108\pm6$&401/359&-64&CPL(VS) \\
\object{130310840} &$0.71\pm0.38$ &$797\pm928$ &395/358 &$71\pm13$ &405/359 &-4 &CPL(P) &$1.15\pm0.02$  &$4560\pm309$&359/283&$76\pm2$&2203/284&-1838&CPL(VS) \\
\object{160818198} &$1.20\pm0.29$ &$131\pm82$ &390/358 &$14.3\pm1.0$ &422/359 &-26 &CPL(VS) &$0.95\pm0.14$  &$81\pm16$&411/358&$14.7\pm0.4$&559/359&-142&CPL(VS) \\
\object{170709334} &$0.58\pm0.25$ &$599\pm426$ &373/355 &$83\pm10$ &385/356 &-6 &CPL(P) &$0.18\pm0.25$  &$245\pm79$&315/355&$76.4\pm5.6$&328/356&-7&CPL(P) \\
\object{170726794} &$0.42\pm0.35$ &$386\pm219$ &350/356 &$92\pm11$ &358/357 &-2 &CPL(NM) & $0.57\pm0.23$ &$320\pm126$&381/356&$57.8\pm4.4$&409/357&-22&CPL(VS) \\
\hline
\enddata
\tablenotetext{a}{The BIC-selected model: a) $0\sim2$, not worth more than a bare mention (NM);
b) $2\sim6$, positive (P); c) $6\sim10$, strong (S); d) $>10$, very strong (VS).}
\label{tab:spectral fitting}
\end{deluxetable}


\clearpage

\begin{figure}
\includegraphics[angle=0,width=0.5\textwidth]{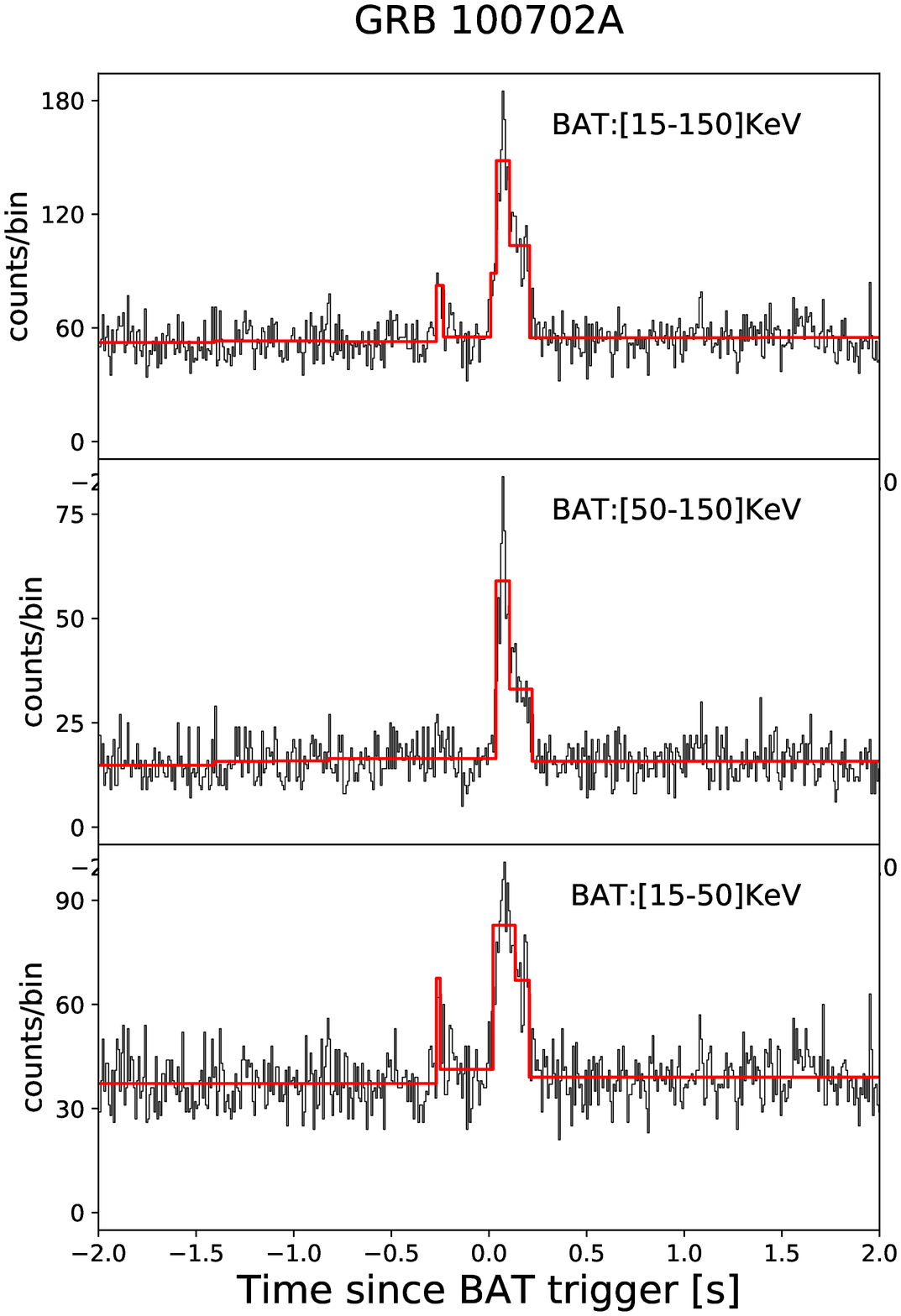}
\center{Fig.1---continued.}
\end{figure}

\clearpage
\begin{figure}
\begin{tabular}{lllll}
\multirow{2}{*}
{\includegraphics[angle=0,width=0.5\textwidth, height=0.32\textheight]{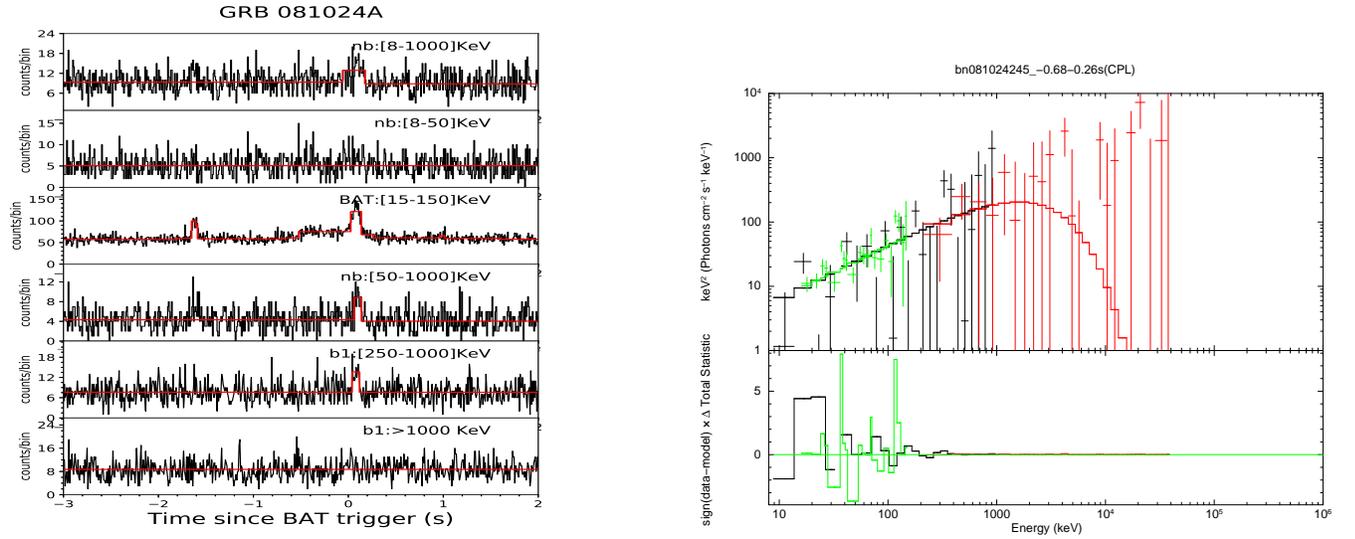}}\\&
\includegraphics[angle=-90,width=0.5\textwidth]{f2a2.eps}\\
\end{tabular}
\caption{{\em Left Panel}: The SGRB light curves with precursors (since BAT trigger) observed by
{\em Swift} and/or {\em Fermi} satellites, which are identified by the BBlocks analysis.
{\em Right Panel}: The better fitting spectra (CPL or BB model) for precursors (top) and main bursts (bottom) (time intervals also labeled since BAT trigger).
The green points, black points, red points, and blue points are data observed with
{\em Swift} BAT, {\em Fermi} NaI, BGO, and LAT, respectively.}
\label{LC_Spec(Swift+Fermi)}
\end{figure}
\clearpage

\begin{figure}
\begin{tabular}{lllll}
\multirow{2}{*}
{\includegraphics[angle=0,width=0.5\textwidth, height=0.63\textheight]{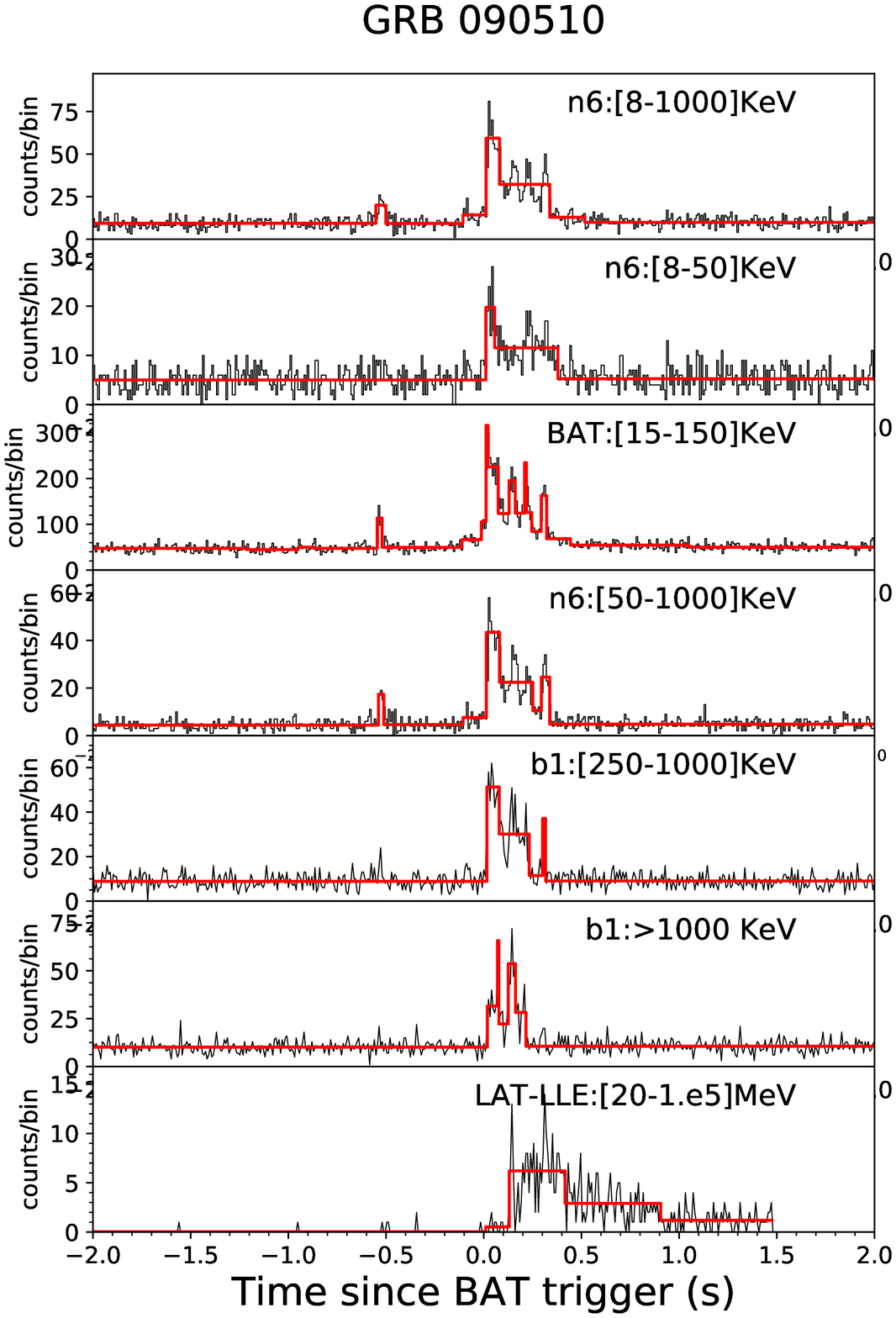}}\\&
\includegraphics[angle=-90,width=0.5\textwidth]{f2b1.eps}\\&
\includegraphics[angle=-90,width=0.5\textwidth]{f2b2.eps}\\
\end{tabular}
\center{Fig.2---continued.}
\end{figure}

\begin{figure}
\begin{tabular}{lllll}
\multirow{2}{*}
{\includegraphics[angle=0,width=0.5\textwidth, height=0.63\textheight]{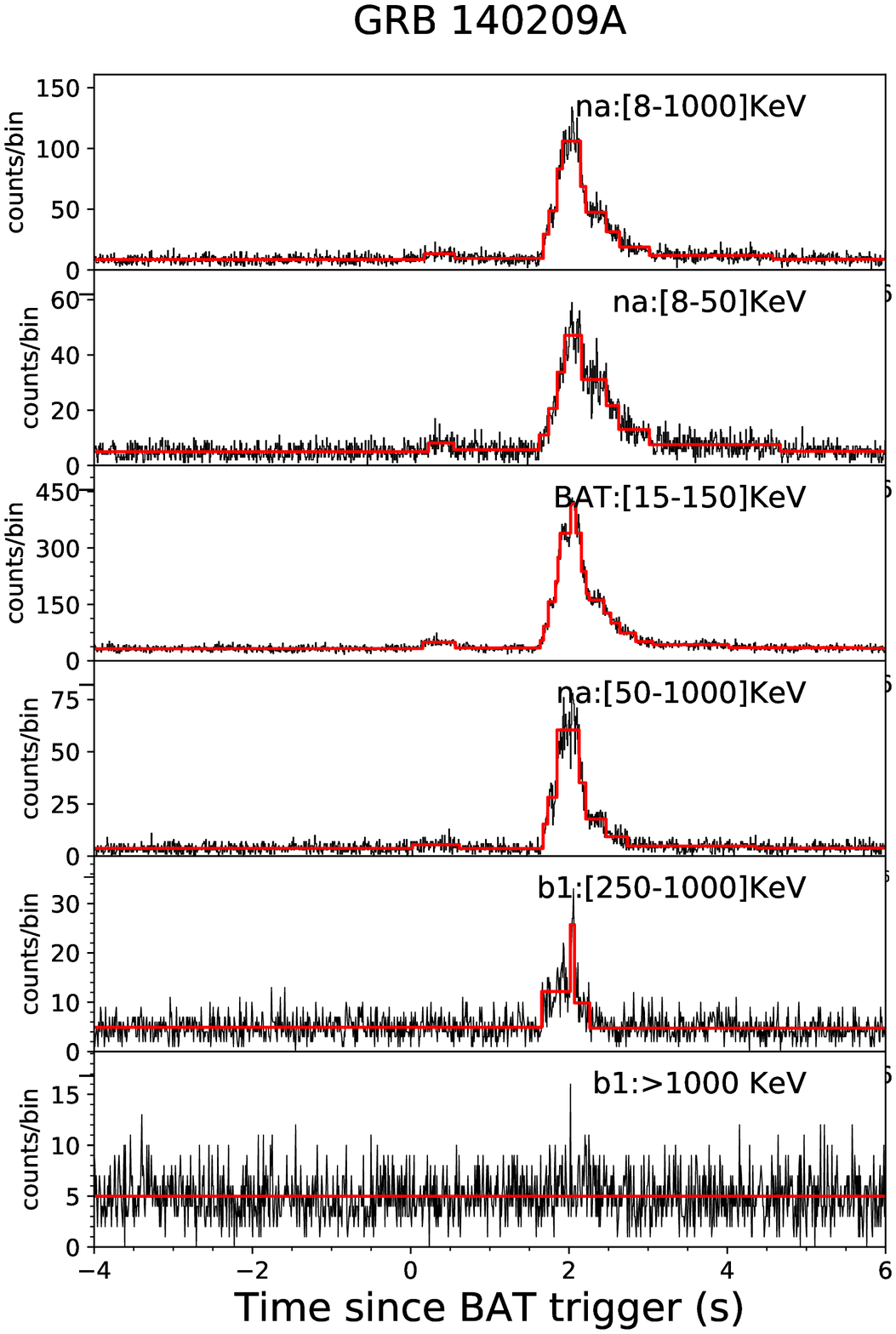}}\\&
\includegraphics[angle=-90,width=0.5\textwidth]{f2c1.eps}\\&
\includegraphics[angle=-90,width=0.5\textwidth]{f2c2.eps}\\
\end{tabular}
\center{Fig.2---continued.}
\end{figure}

\clearpage
\begin{figure}
\begin{tabular}{lllll}
\multirow{2}{*}
{\includegraphics[angle=0,width=0.5\textwidth, height=0.32\textheight]{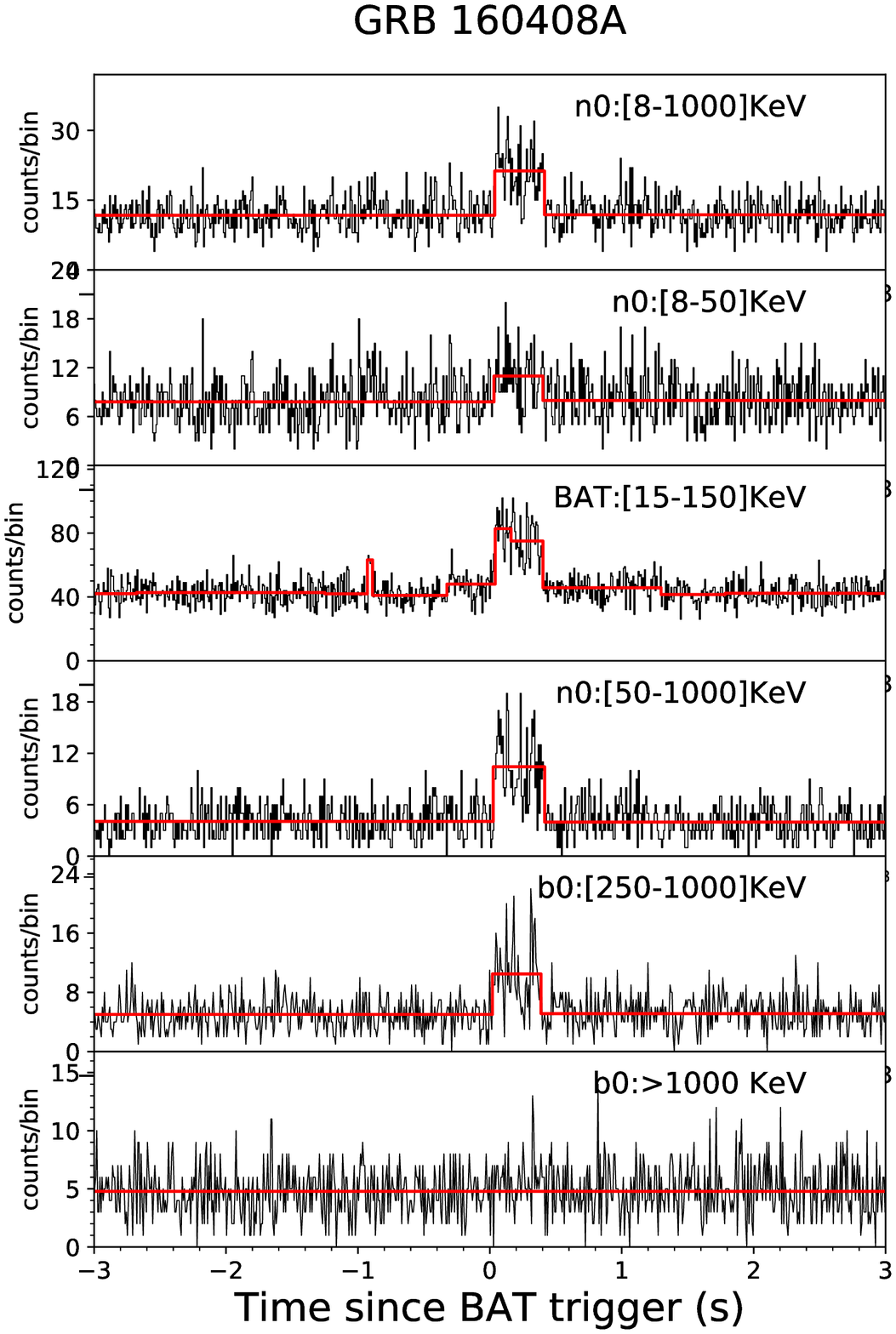}}\\&
\includegraphics[angle=-90,width=0.5\textwidth]{f2d2.eps}\\
\end{tabular}
\center{Fig.2---continued.}
\end{figure}
\clearpage

\begin{figure}
\begin{tabular}{lllll}
\multirow{2}{*}
{\includegraphics[angle=0,width=0.5\textwidth, height=0.63\textheight]{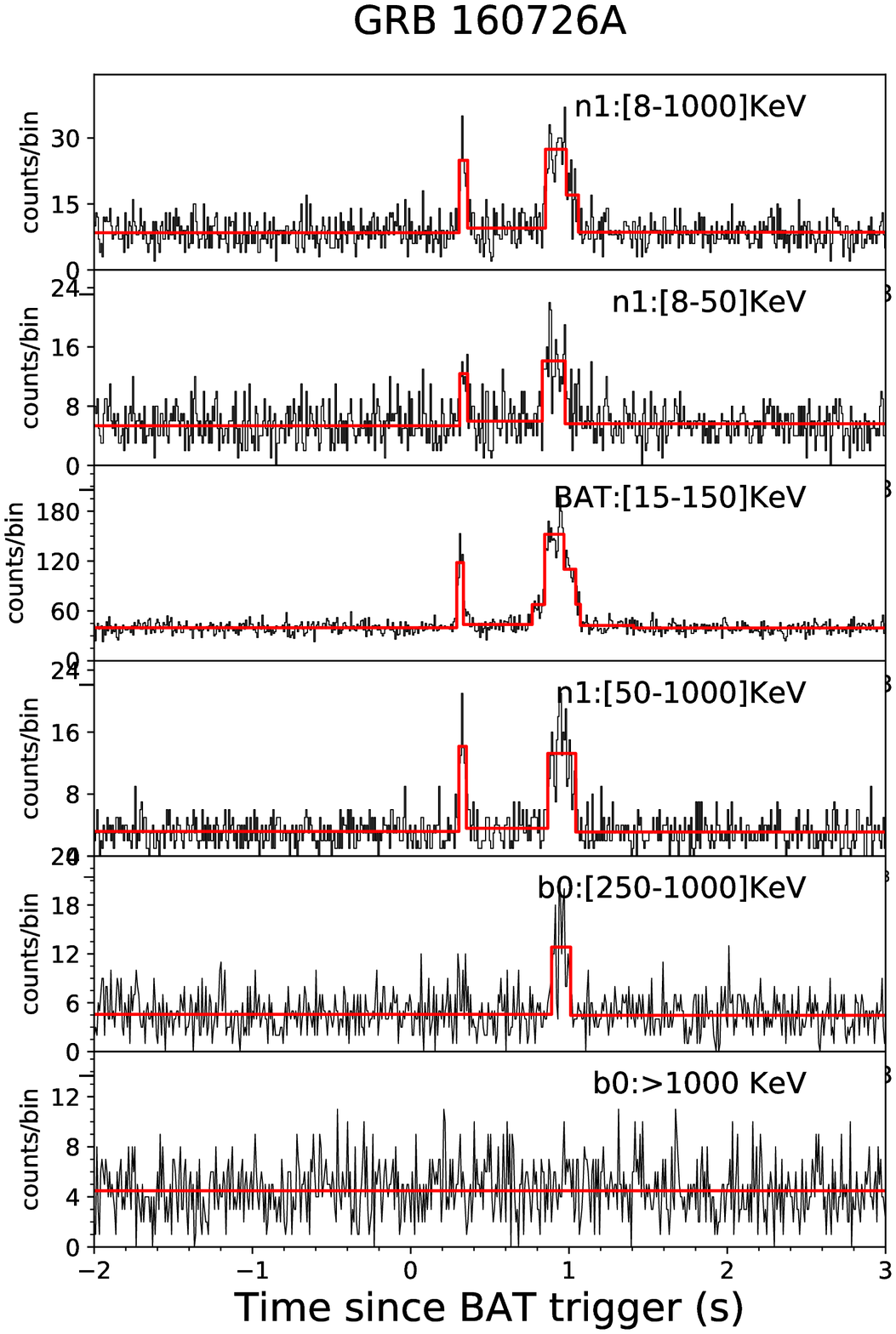}}\\&
\includegraphics[angle=-90,width=0.5\textwidth]{f2e1.eps}\\&
\includegraphics[angle=-90,width=0.5\textwidth]{f2e2.eps}\\
\end{tabular}
\center{Fig.2---continued.}
\end{figure}

\begin{figure}
\begin{tabular}{lllll}
\multirow{2}{*}
{\includegraphics[angle=0,width=0.5\textwidth, height=0.63\textheight]{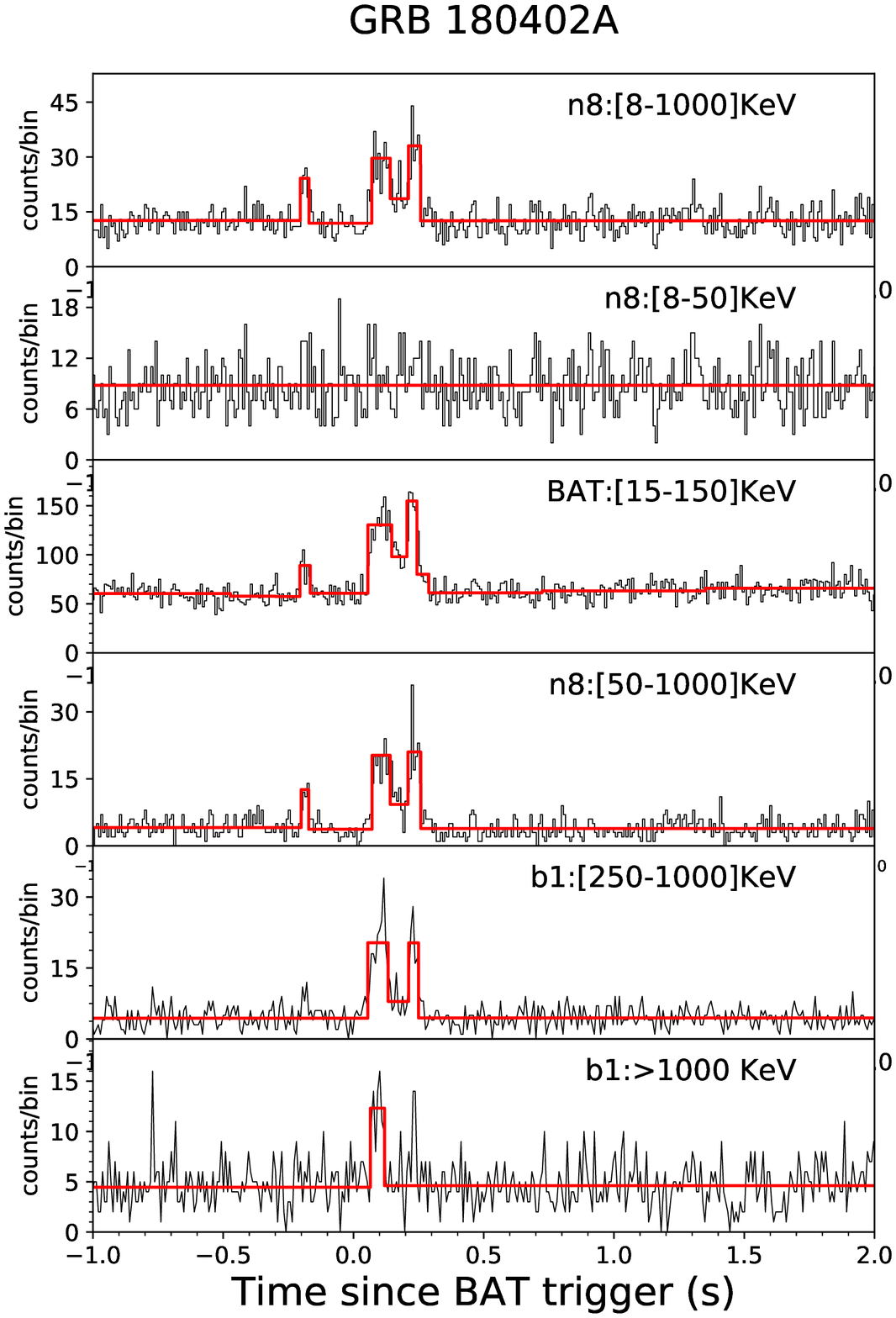}}\\&
\includegraphics[angle=-90,width=0.5\textwidth]{f2f1.eps}\\&
\includegraphics[angle=-90,width=0.5\textwidth]{f2f2.eps}\\
\end{tabular}
\center{Fig.2---continued.}
\end{figure}

\clearpage
\begin{figure}
\begin{tabular}{lllll}
\multirow{2}{*}
{\includegraphics[angle=0,width=0.5\textwidth, height=0.63\textheight]{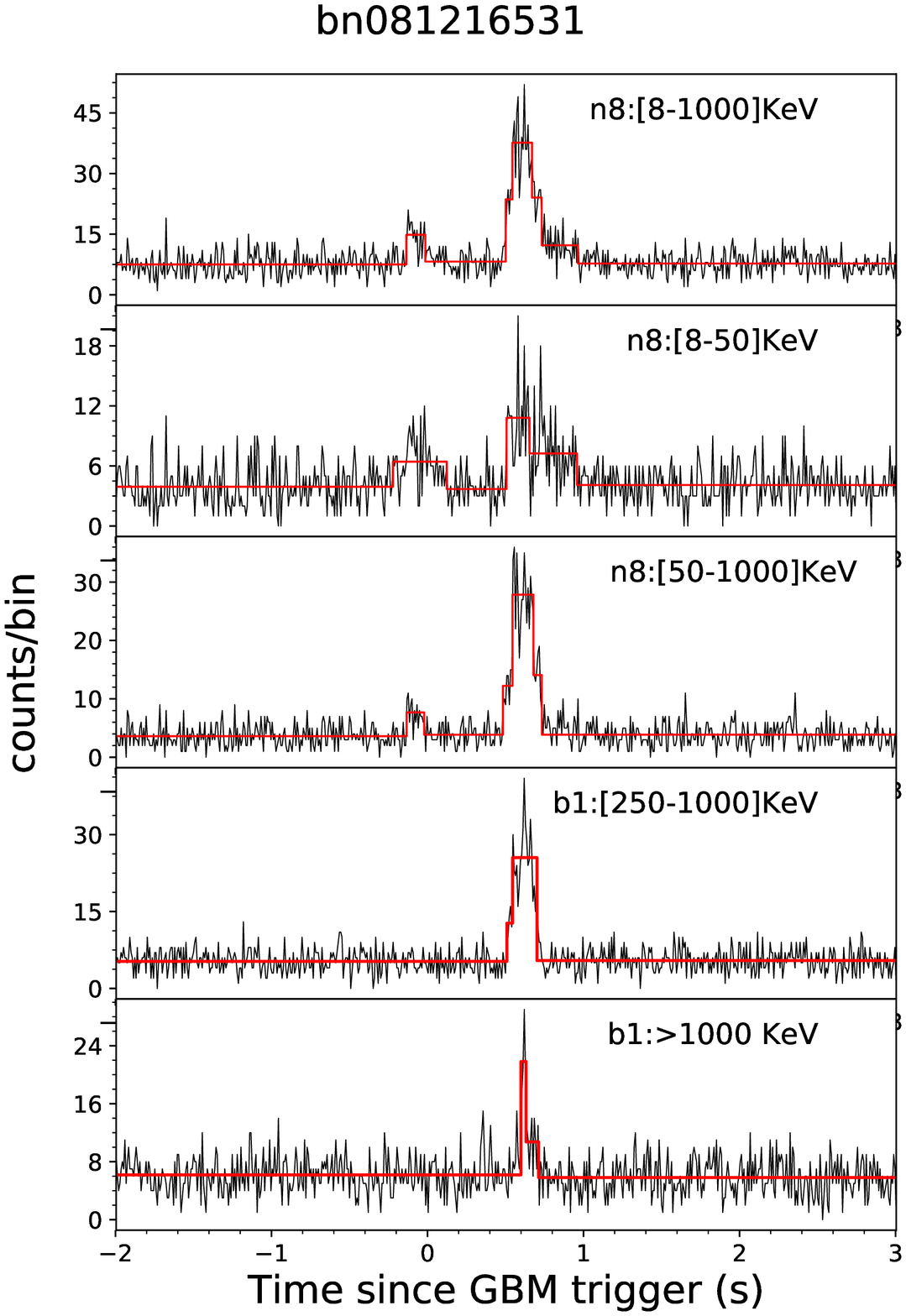}}\\&
\includegraphics[angle=-90,width=0.5\textwidth]{f3a1.eps}\\&
\includegraphics[angle=-90,width=0.5\textwidth]{f3a2.eps}\\
\end{tabular}
\center\caption{{\em Left Panel}: The SGRB light curves with precursors
observed by only {\em Fermi} satellite, which are identified by the BBlocks analysis.
{\em Right Panel}: The better fitting spectra (CPL or BB model) for precursors (top) and main bursts (bottom).
The red points and black points, and green points are data observed with {\em Fermi} NaI, and BGO, respectively.}
\label{LC_Spec(Fermi)}
\end{figure}

\begin{figure}
\begin{tabular}{lllll}
\multirow{2}{*}
{\includegraphics[angle=0,width=0.5\textwidth, height=0.63\textheight]{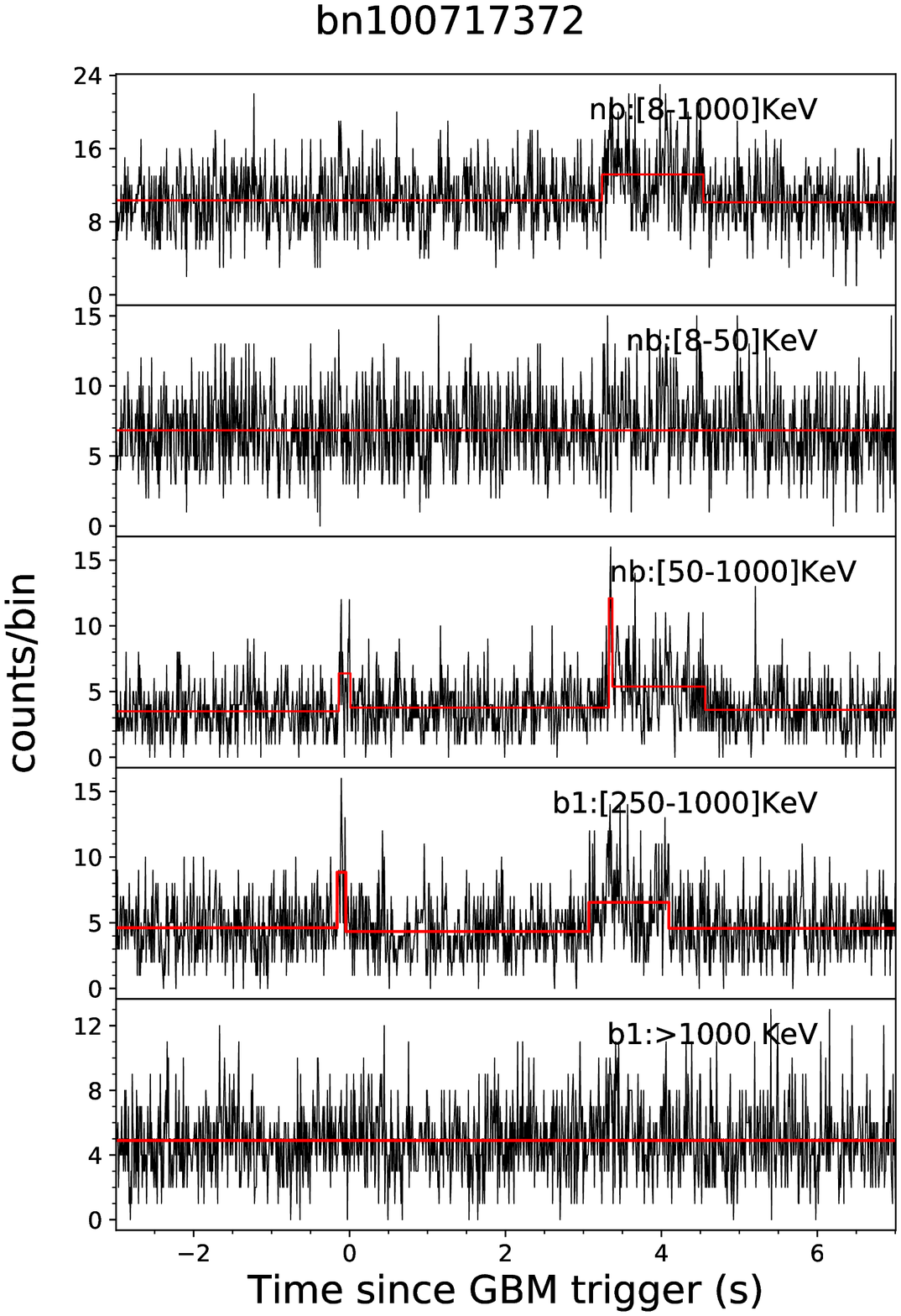}}\\&
\includegraphics[angle=-90,width=0.5\textwidth]{f3b1.eps}\\&
\includegraphics[angle=-90,width=0.5\textwidth]{f3b2.eps}\\
\end{tabular}
\center{Fig.3---continued.}
\end{figure}

\begin{figure}
\begin{tabular}{lllll}
\multirow{2}{*}
{\includegraphics[angle=0,width=0.5\textwidth, height=0.63\textheight]{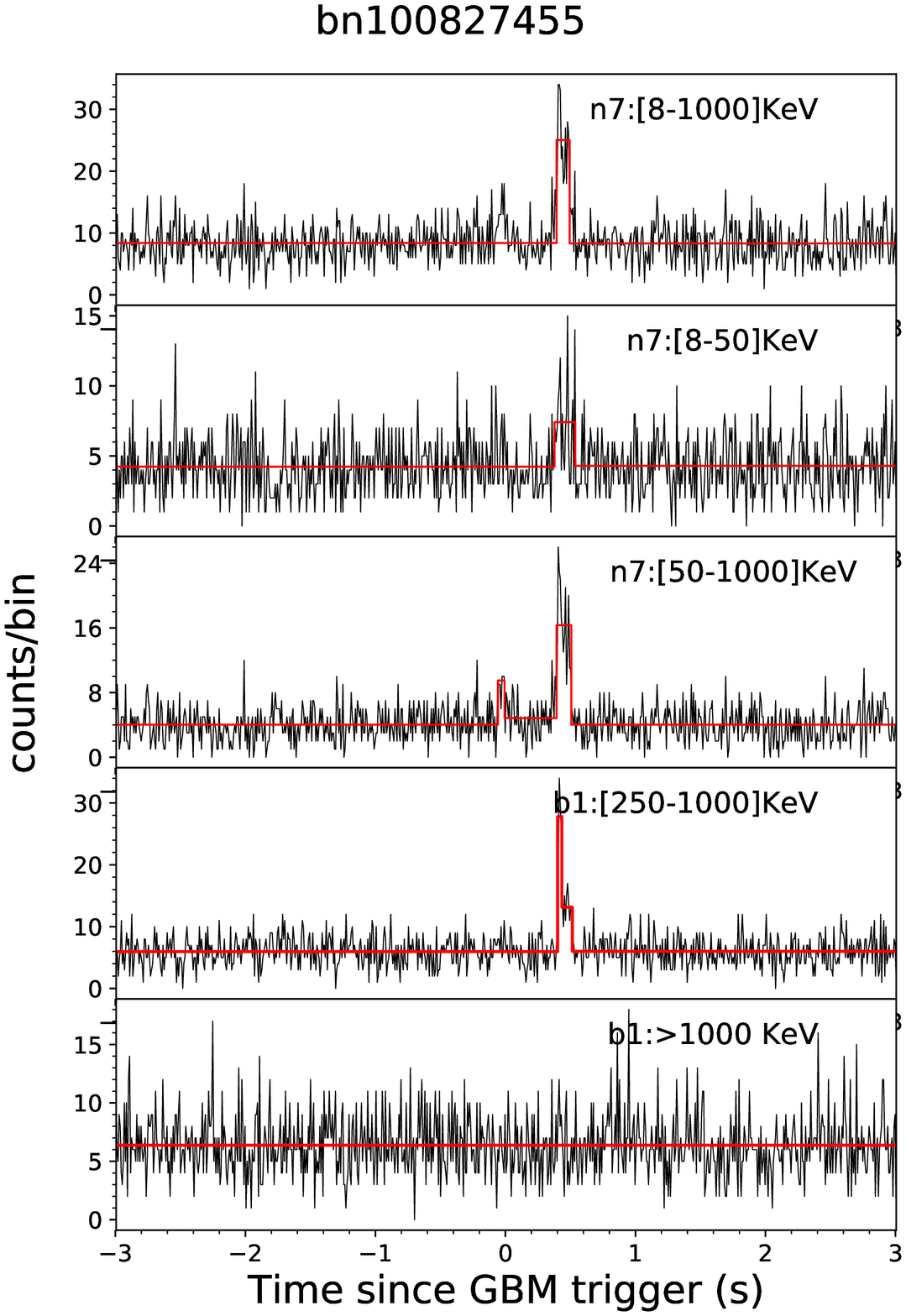}}\\&
\includegraphics[angle=-90,width=0.5\textwidth]{f3c1.eps}\\&
\includegraphics[angle=-90,width=0.5\textwidth]{f3c2.eps}\\
\end{tabular}
\center{Fig.3---continued.}
\end{figure}

\begin{figure}
\begin{tabular}{lllll}
\multirow{2}{*}
{\includegraphics[angle=0,width=0.5\textwidth, height=0.63\textheight]{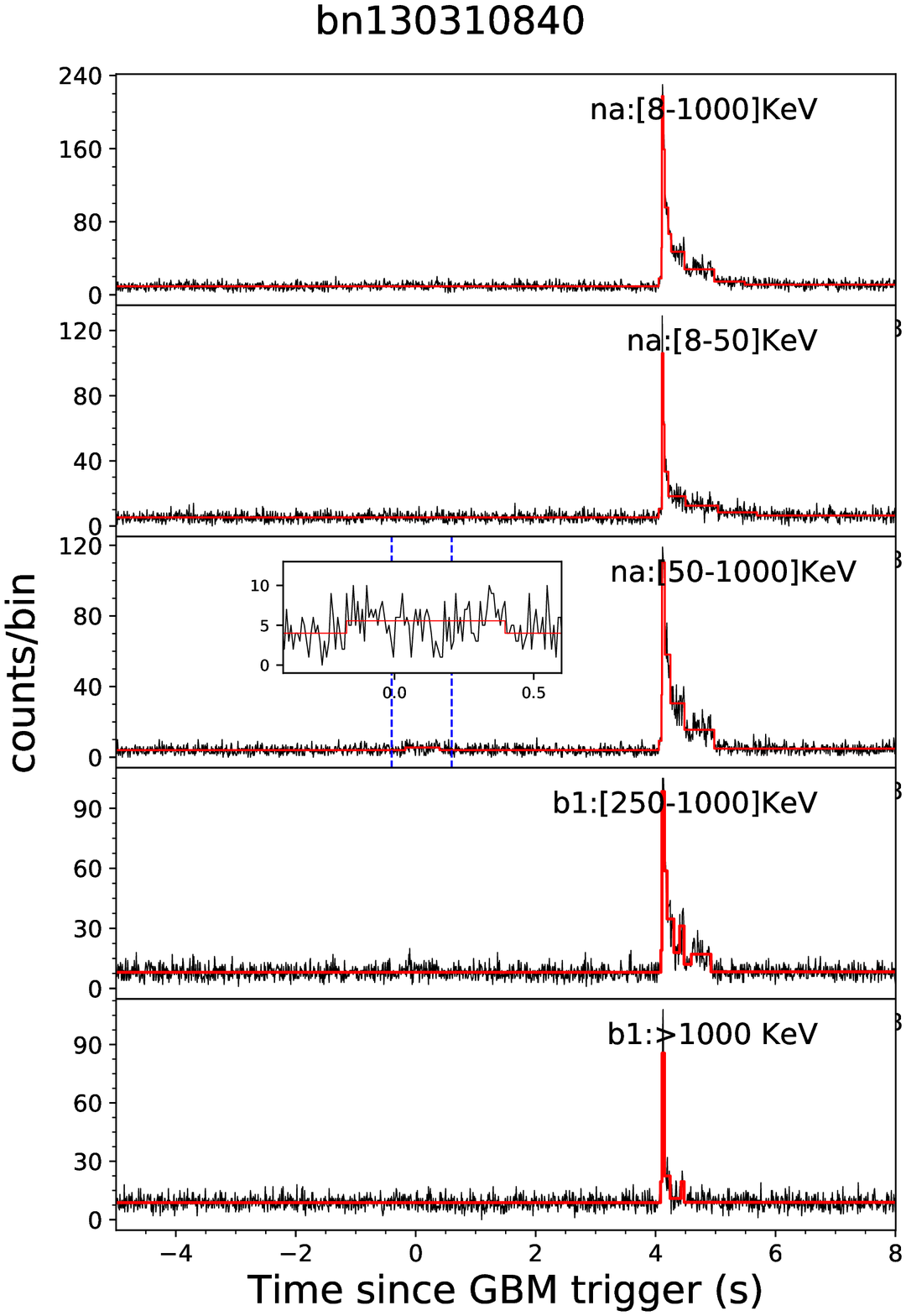}}\\&
\includegraphics[angle=-90,width=0.5\textwidth]{f3d1.eps}\\&
\includegraphics[angle=-90,width=0.5\textwidth]{f3d2.eps}\\
\end{tabular}
\center{Fig.3---continued.}
\end{figure}

\begin{figure}
\begin{tabular}{lllll}
\multirow{2}{*}
{\includegraphics[angle=0,width=0.5\textwidth, height=0.63\textheight]{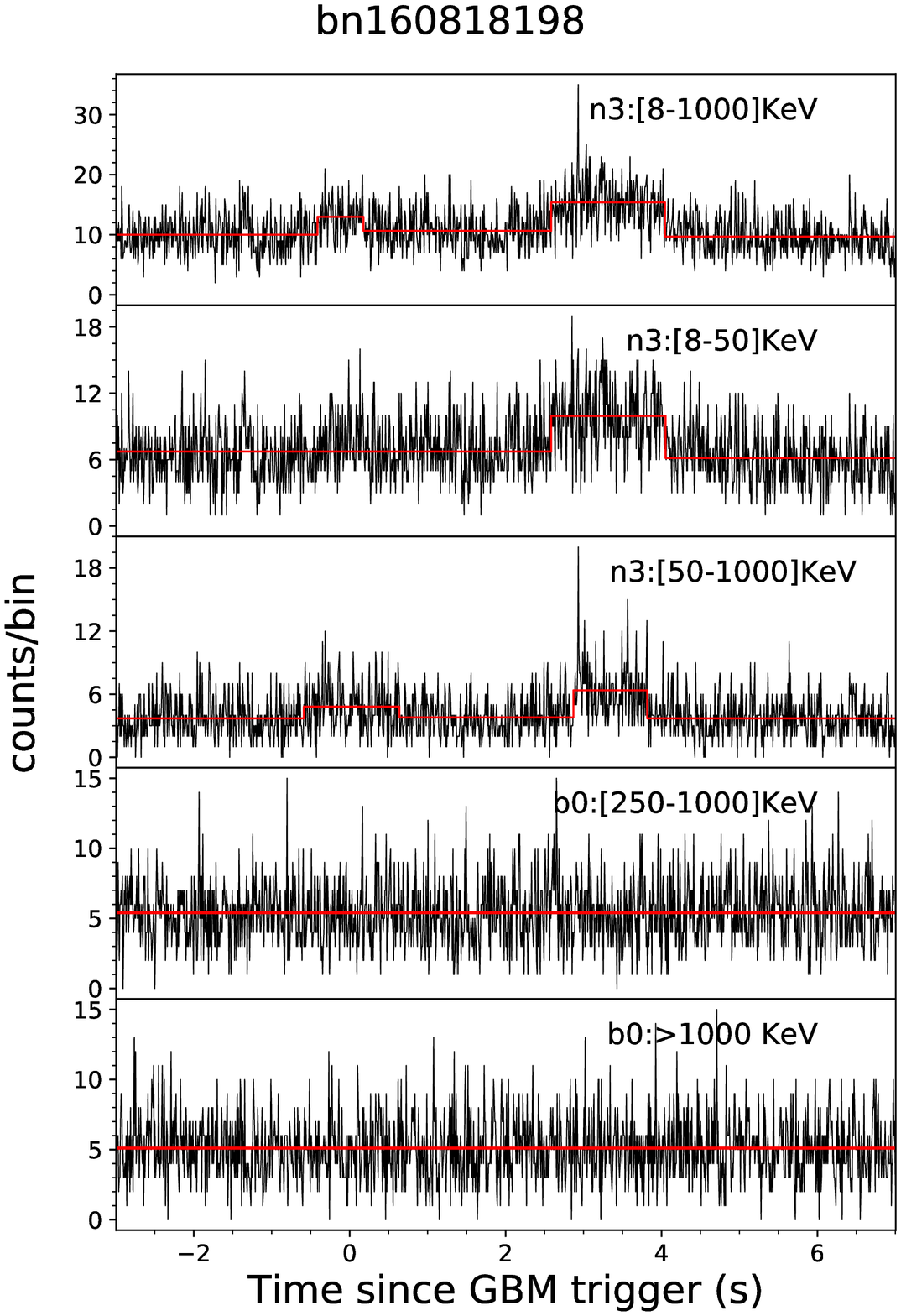}}\\&
\includegraphics[angle=-90,width=0.5\textwidth]{f3e1.eps}\\&
\includegraphics[angle=-90,width=0.5\textwidth]{f3e2.eps}\\
\end{tabular}
\center{Fig.3---continued.}
\end{figure}

\begin{figure}
\begin{tabular}{lllll}
\multirow{2}{*}
{\includegraphics[angle=0,width=0.5\textwidth, height=0.65\textheight]{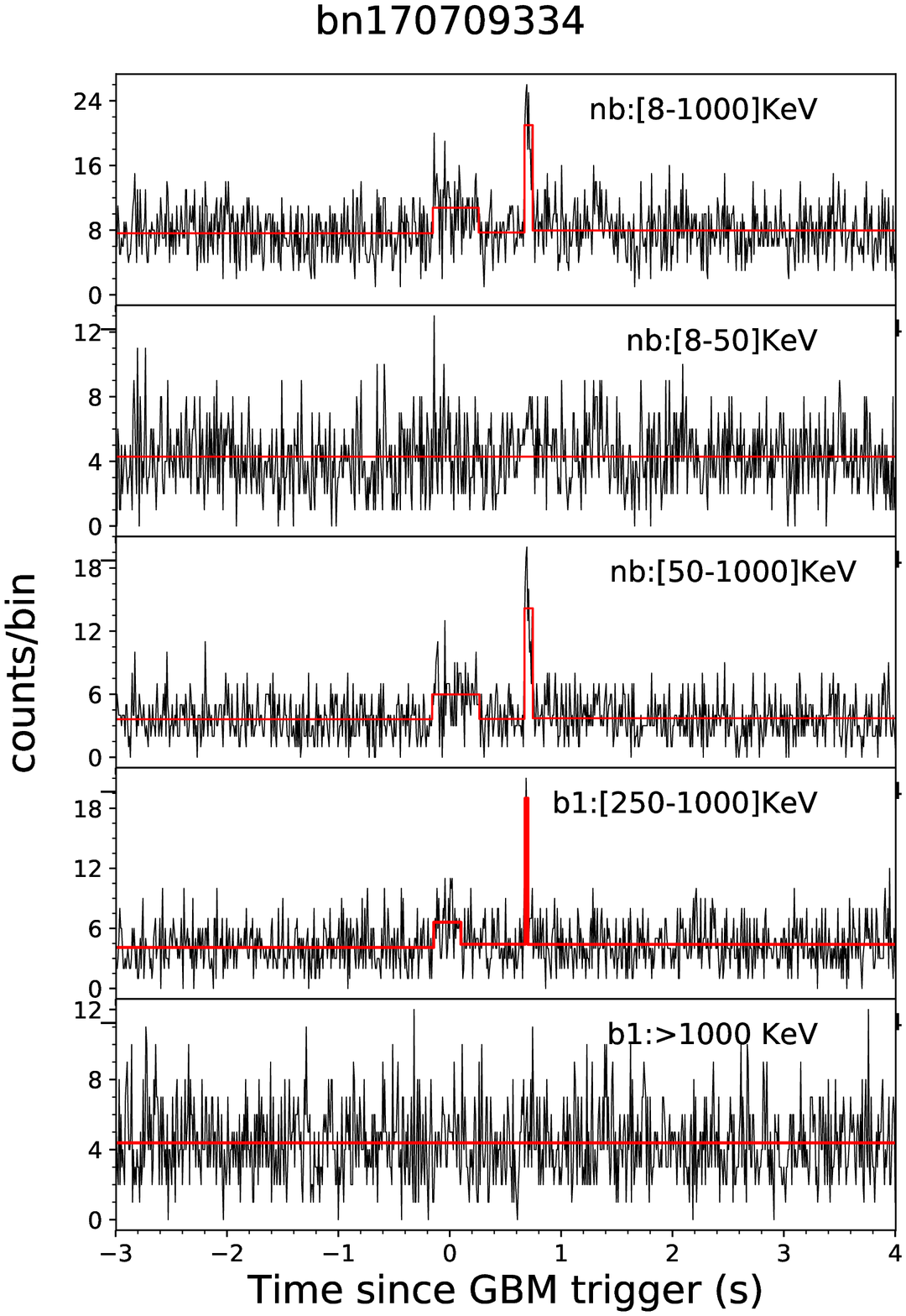}}\\&
\includegraphics[angle=-90,width=0.5\textwidth]{f3f1.eps}\\&
\includegraphics[angle=-90,width=0.5\textwidth]{f3f2.eps}\\
\end{tabular}
\center{Fig.3---continued.}
\end{figure}

\begin{figure}
\begin{tabular}{lllll}
\multirow{2}{*}
{\includegraphics[angle=0,width=0.5\textwidth, height=0.63\textheight]{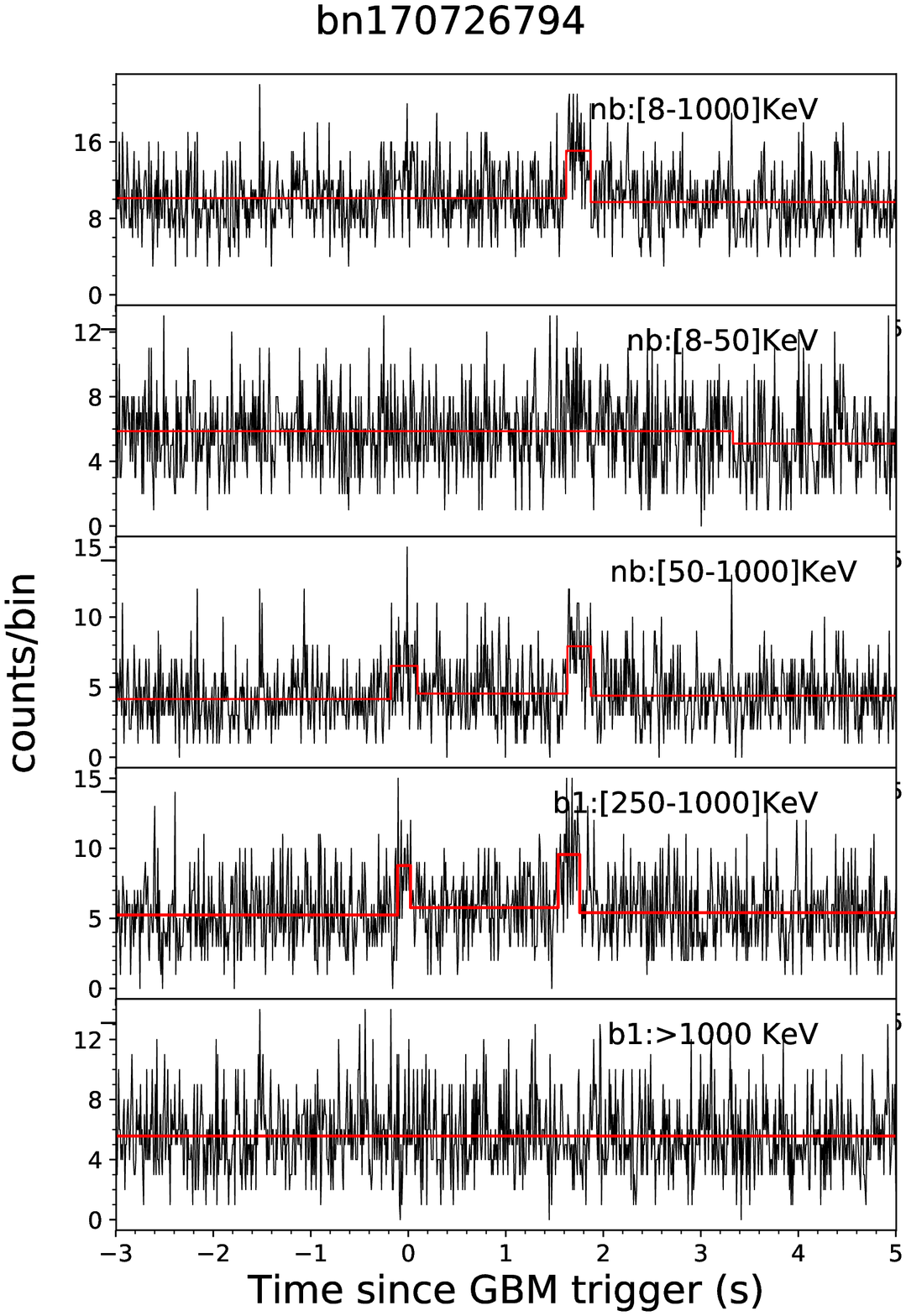}}\\&
\includegraphics[angle=-90,width=0.5\textwidth]{f3g1.eps}\\&
\includegraphics[angle=-90,width=0.5\textwidth]{f3g2.eps}\\
\end{tabular}
\center{Fig.3---continued.}
\end{figure}

\clearpage
\begin{figure}
\includegraphics[angle=0,width=0.5\textwidth]{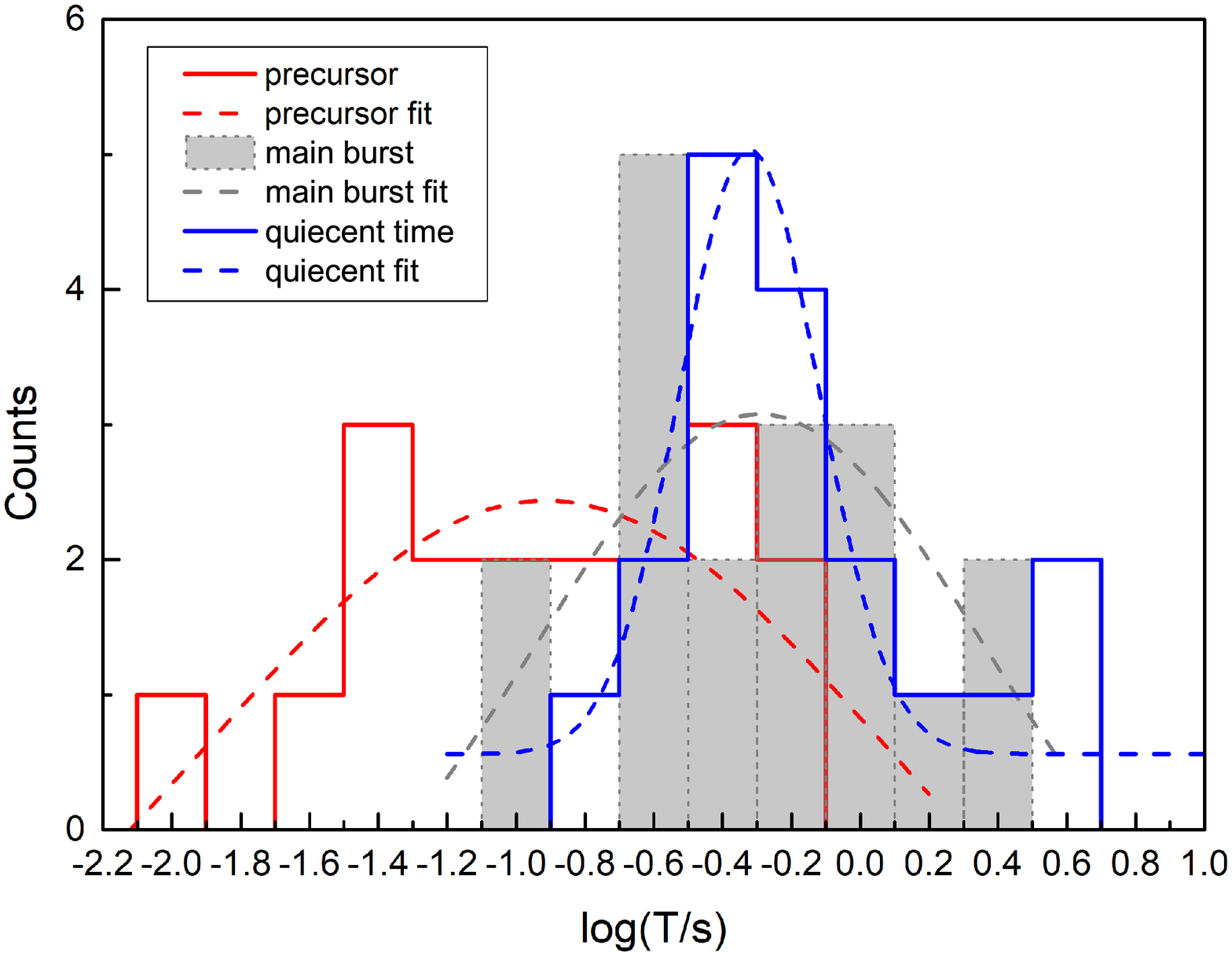}
\includegraphics[angle=0,width=0.5\textwidth]{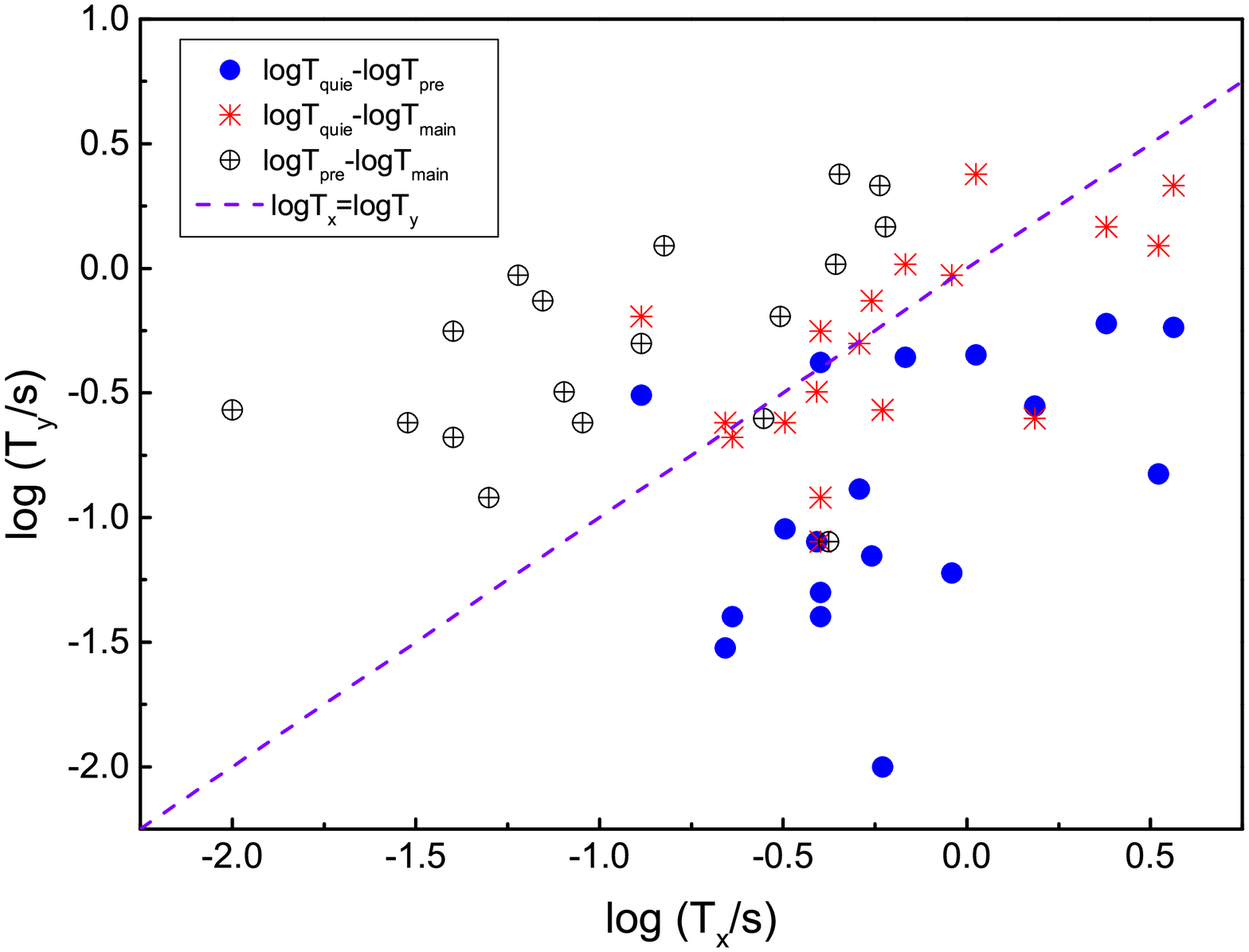}
\caption{{\em Left Panel}: The distributions of precursor duration,
main burst duration, and quiescent time. {\em Right Panel}: The possible relations among these three quantities.}
\label{Time}
\end{figure}

\begin{figure}
\includegraphics[angle=0,width=0.5\textwidth]{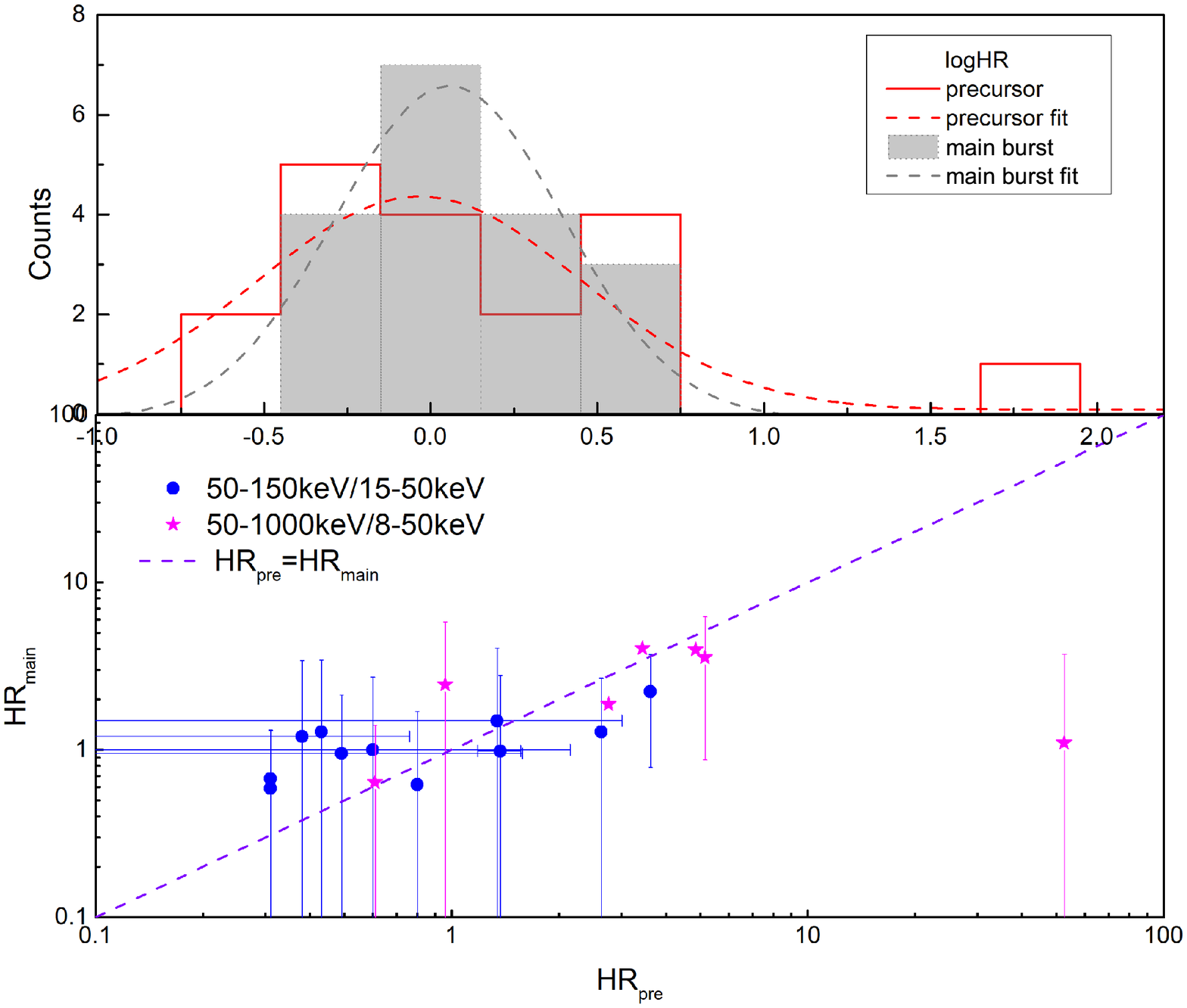}
\includegraphics[angle=0,width=0.5\textwidth]{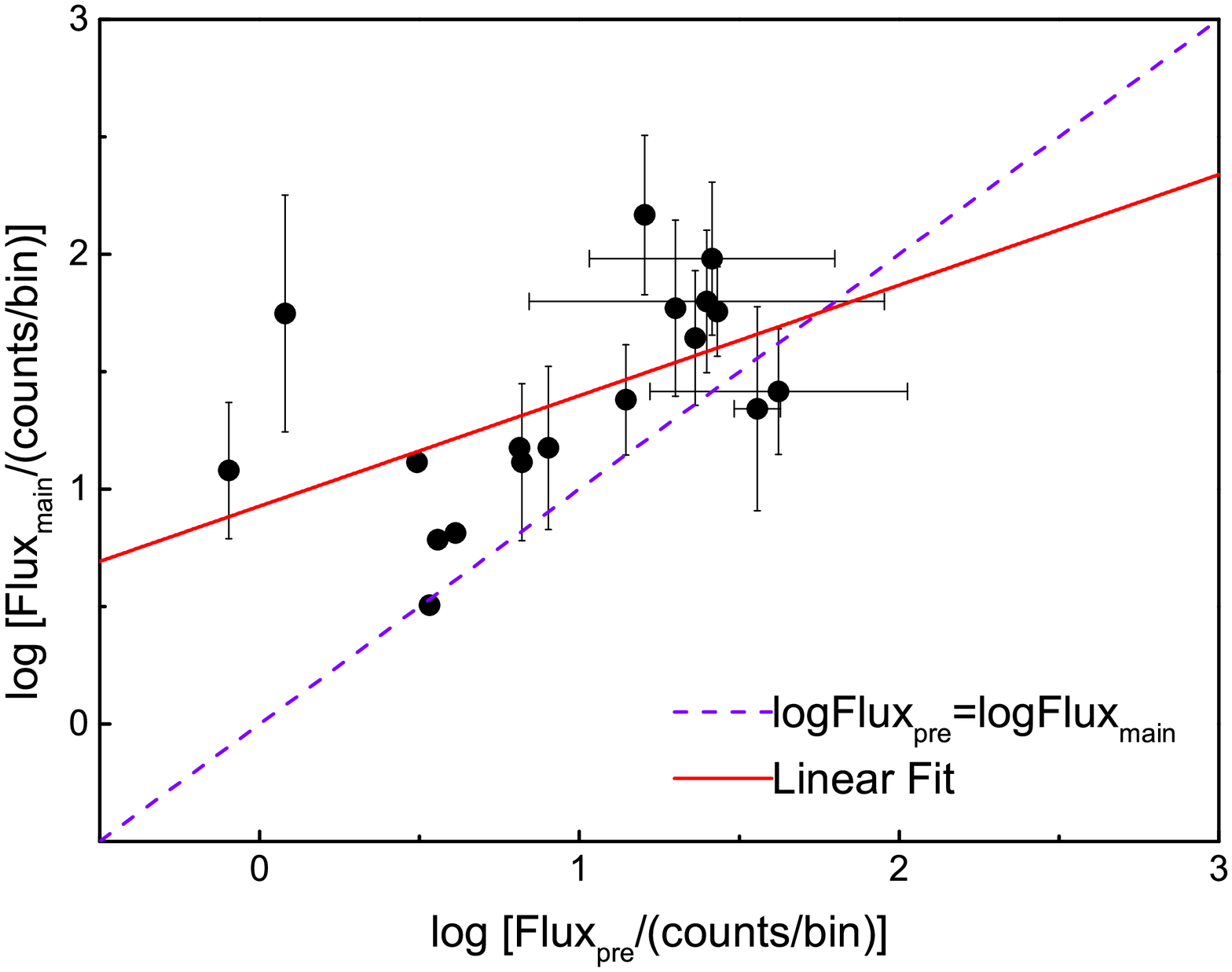}
\caption{{\em Left Panel}: The comparison of hardness ratio (HR) between the precursors and main bursts. For SGRBs observed by only {\em Swift} and by both {\em Swift} and {\em Fermi}, the HR is defined as the ratio of the count rates in 50-150 keV over 15-50 keV. The HR in SGRBs observed by only {\em Fermi}, is defined as the ratio of count rate in 50-1000 keV over 8-50 keV. {\em Right Panel}: The average flux for each precursor and main burst.}
\label{hardness}
\end{figure}

\begin{figure}
\includegraphics[angle=0,width=0.5\textwidth]{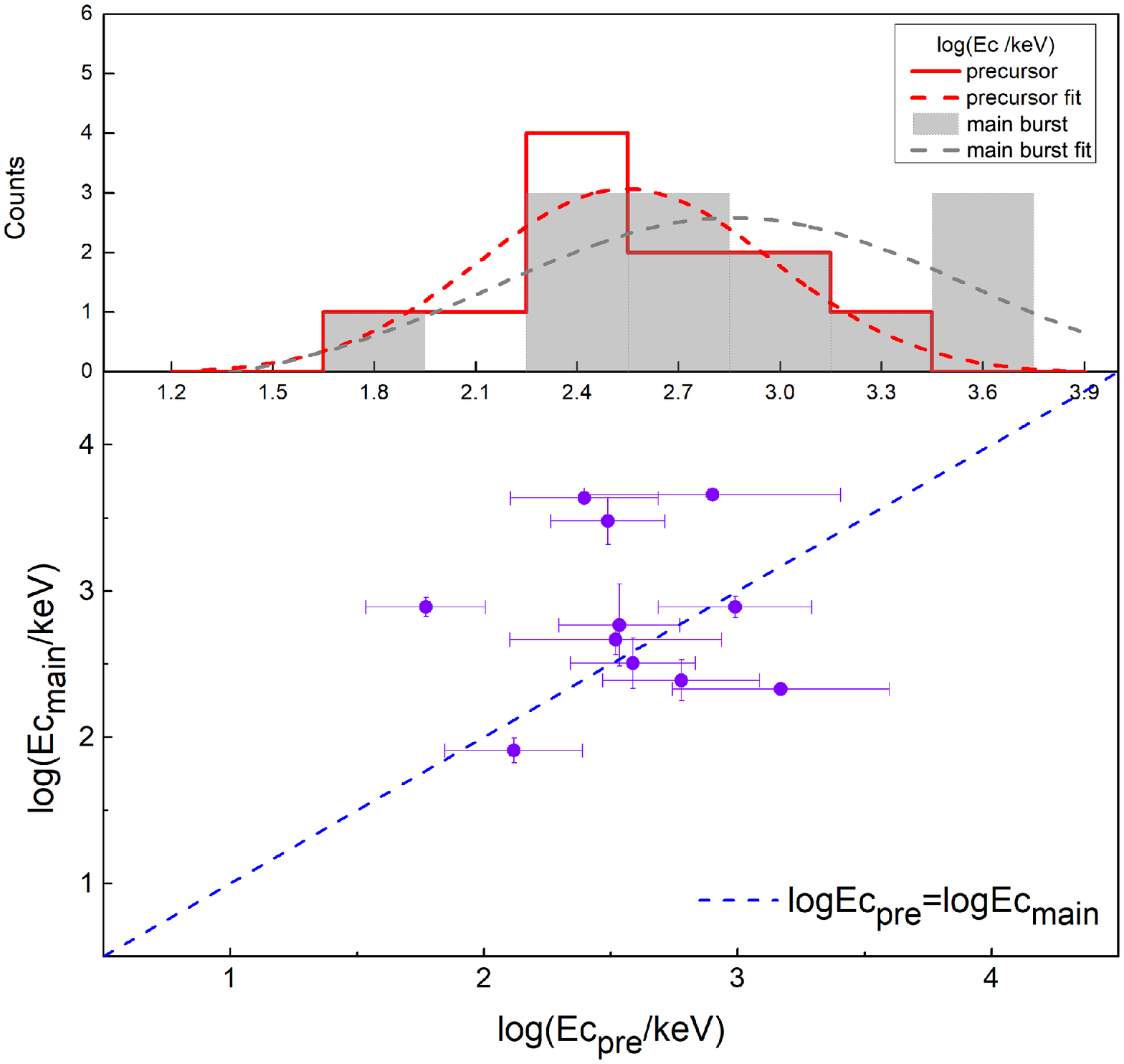}
\includegraphics[angle=0,width=0.5\textwidth]{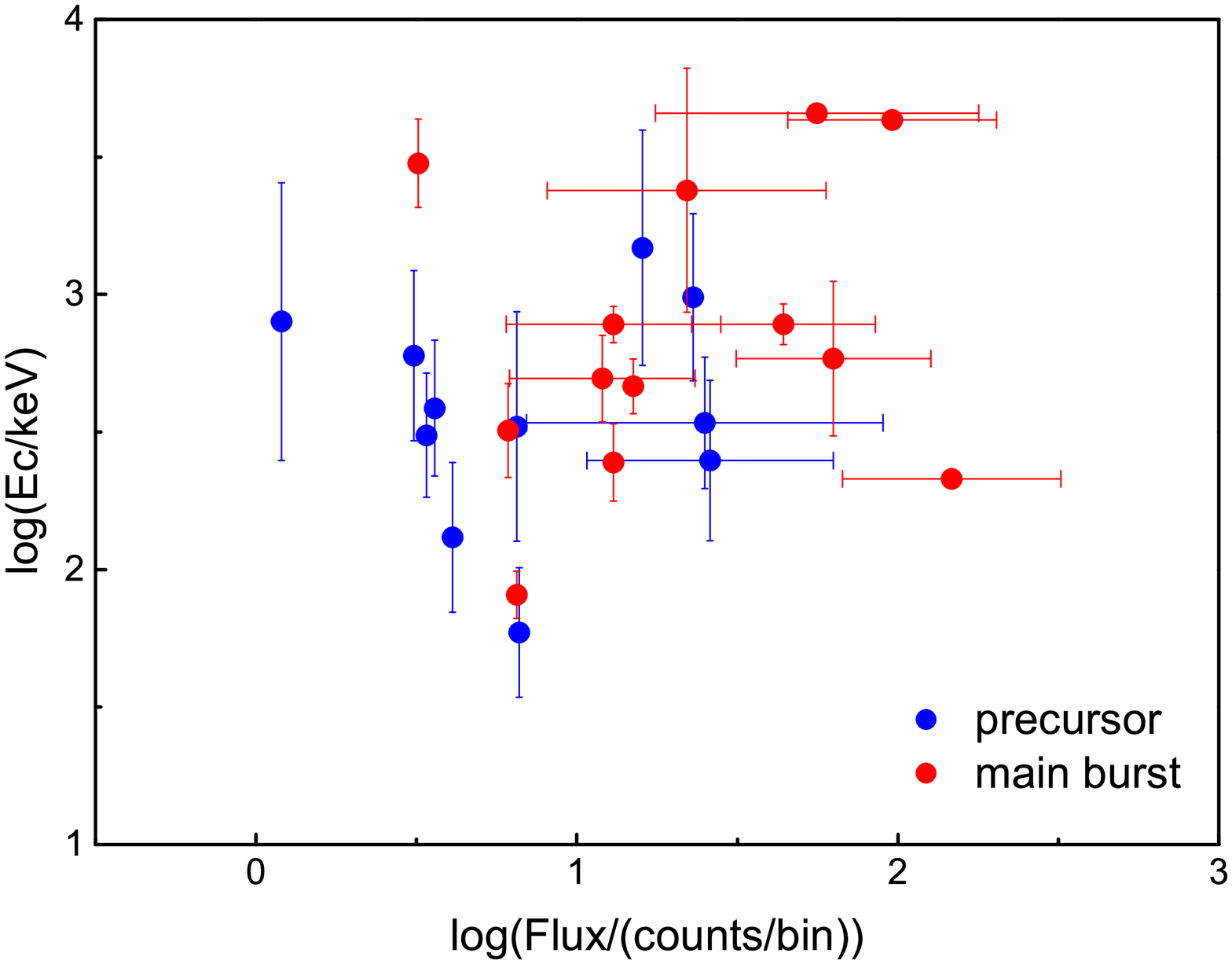}
\caption{{\em Left Panel}: The distributions of cutoff energy $E_{\rm c}$ for the precursors and main bursts,
and the cutoff energy of the precursors as a function of that of the main bursts. {\em Right Panel}: The cutoff energy as a function of the corresponding average flux for the precursors and main bursts.}
\label{spec_Ec}
\end{figure}


\end{document}